\documentclass[iop,revtex4,twocolappendix]{emulateapj}

\newcommand{\Kepler}{{\it Kepler}}
\newcommand{\Ktwo}{{\it K2}}

\newcommand{\vespa}{{\texttt{vespa}}}
\usepackage{color}

\newcommand{\ntargets}{$208,423$}% C4-C10 LC targets + 59,174 (see Vanderburg 2016a abstract). Note: we only went down to V=13.

\newcommand{\ntces}{${\sim} 30,000$} % ~number of TCEs found

\newcommand{\ncands}{${\sim} 1,000$}% ~number of candidates (passed triage + vetting)

\newcommand{\ntrescands}{$275$}% number of candidates to pass vetting
\newcommand{\ntrescandsys}{$233$}% number of systems hosting final candidates
\newcommand{\nspeckleaosys}{$186$}% number of systems with speckle or AO data

\newcommand{\nunvalidated}{$126$}% number of unvalidated candidates

\newcommand{\nplanets}{$149$}% number of validated planets
\newcommand{\nsys}{$111$}% number of validated planet systems

\newcommand{\nalreadyplanets}{$53$}% number of planets already labeled "validated planet"
\newcommand{\nalreadycand}{$39$}% number of planets already labeled "candidate"
\newcommand{\nnew}{$56$}% number of planets not yet found
\newcommand{\nfp}{$1$}% number of planets already labeled "false positive"

\usepackage{makecell}
\usepackage{hhline}

\usepackage{lscape}
\usepackage{amsmath}

\slugcomment{}

\shorttitle{New Candidates and Planets Transiting Bright Stars}
\shortauthors{Mayo et al.}

\begin{document}


\title{275 Candidates and 149 Validated Planets Orbiting Bright Stars in \Ktwo\ Campaigns 0-10}


\author{Andrew W. Mayo\altaffilmark{1,2,3,$\dagger$,$\ddagger$,$\star$}, Andrew Vanderburg\altaffilmark{4,3,$\ddagger$, $\Box$}, David W. Latham\altaffilmark{3}, Allyson Bieryla\altaffilmark{3}, Timothy D. Morton\altaffilmark{5}, Lars A. Buchhave\altaffilmark{1,2}, Courtney D. Dressing\altaffilmark{6,7,$\Box$}, Charles Beichman\altaffilmark{8}, Perry Berlind\altaffilmark{3}, Michael L. Calkins\altaffilmark{3}, David R. Ciardi\altaffilmark{8}, Ian J. M. Crossfield\altaffilmark{9}, Gilbert A. Esquerdo\altaffilmark{3}, Mark E. Everett\altaffilmark{10}, Erica J. Gonzales\altaffilmark{11, $\ddagger$}, Lea A. Hirsch\altaffilmark{6}, Elliott P. Horch\altaffilmark{12}, Andrew W. Howard\altaffilmark{13}, Steve B. Howell\altaffilmark{14}, John Livingston\altaffilmark{15}, Rahul Patel\altaffilmark{16} , Erik A. Petigura\altaffilmark{13, $\dagger\dagger$}, Joshua E. Schlieder\altaffilmark{17}, Nicholas J. Scott\altaffilmark{14}, Clea F. Schumer\altaffilmark{3}, Evan Sinukoff\altaffilmark{13,18}, Johanna Teske\altaffilmark{19,$\ddagger\ddagger$}, Jennifer G. Winters\altaffilmark{3}}

%% are identified by the \altaffilmark after each name. Specify alternate
\altaffiltext{$\dagger$}{\texttt{awm@space.dtu.dk}}
\altaffiltext{$\ddagger$}{National Science Foundation Graduate Research Fellow}
\altaffiltext{$\star$}{Fulbright Fellow}
\altaffiltext{$\Box$}{NASA Sagan Fellow}
\altaffiltext{$\dagger\dagger$}{NASA Hubble Fellow}
\altaffiltext{$\ddagger\ddagger$}{Carnegie Origins Fellow, jointly appointed by Carnegie DTM \& Carnegie Observatories}
\altaffiltext{1}{DTU Space, National Space Institute, Technical University of Denmark, Elektrovej 327, DK-2800 Lyngby, Denmark}
\altaffiltext{2}{Centre for Star and Planet Formation, Natural History Museum of Denmark \& Niels Bohr Institute, University of Copenhagen, \O ster Voldgade 5-7, DK-1350 Copenhagen K., Denmark}
\altaffiltext{3}{Harvard--Smithsonian Center for Astrophysics, 60 Garden Street, Cambridge, MA 02138, USA}
\altaffiltext{4}{Department of Astronomy, University of Texas at Austin, Austin, TX 76712, USA}
\altaffiltext{5}{Department of Astrophysical Sciences, 4 Ivy Lane, Peyton Hall, Princeton University, Princeton, NJ 08544, USA}
\altaffiltext{6}{Astronomy Department, University of California, Berkeley, CA 94720, USA}
\altaffiltext{7}{Division of Geological \& Planetary Sciences, California Institute of Technology, 1200 East California Boulevard MC 150-21, Pasadena, CA 91125, USA}
\altaffiltext{8}{NASA Exoplanet Science Institute, California Institute of Technology, Pasadena, CA 91125, USA}
\altaffiltext{9}{Department of Physics and Kavli Institute for Astrophysics and Space Research, Massachusetts Institute of Technology, Cambridge, MA 02139, USA}
\altaffiltext{10}{National Optical Astronomy Observatory, 950 North Cherry Avenue, Tucson, AZ 85719, USA}
\altaffiltext{11}{Department of Astronomy and Astrophysics, University of California, Santa Cruz, CA 95064, USA}
\altaffiltext{12}{Department of Physics, Southern Connecticut State University, 501 Crescent Street, New Haven, CT 06515, USA}
\altaffiltext{13}{Cahill Center for Astrophysics, California Institute of Technology, Pasadena, CA 91125, USA}
\altaffiltext{14}{Space Science and Astrobiology Division, NASA Ames Research Center, Moffett Field, CA 94035, USA}
\altaffiltext{15}{Department of Astronomy, The University of Tokyo, 7-3-1 Hongo, Bunkyo-ku, Tokyo 113-0033, Japan}
\altaffiltext{16}{Infrared Processing and Analysis Center, California Institute of Technology, Pasadena, CA 91125, USA}
\altaffiltext{17}{NASA Goddard Space Flight Center, Greenbelt, MD 20771, USA}
\altaffiltext{18}{Institute for Astronomy, University of Hawai`i at M\={a}noa, Honolulu, HI 96822, USA}
\altaffiltext{19}{Carnegie Observatories, 813 Santa Barbara Street, Pasadena, CA 91101, USA}


\begin{abstract}
Since 2014, NASA's \Ktwo\ mission has observed large portions of the ecliptic plane in search of transiting planets and has detected hundreds of planet candidates. With observations planned until at least early 2018, \Ktwo\ will continue to identify more planet candidates. We present here \ntrescands\ planet candidates observed during Campaigns 0-10 of the \Ktwo\ mission that are orbiting stars brighter than 13 mag (in Kepler band) and for which we have obtained high-resolution spectra $(R = 44,000)$. These candidates are analyzed using the \vespa\ package \citep{morton2012, morton2015b} in order to calculate their false-positive probabilities (FPP). We find that \nplanets\ candidates are validated with an FPP lower than 0.1\%, \nalreadycand\ of which were previously only candidates and \nnew\ of which were previously undetected. The processes of data reduction, candidate identification, and statistical validation are described, and the demographics of the candidates and newly validated planets are explored. We show tentative evidence of a gap in the planet radius distribution of our candidate sample. Comparing our sample to the \Kepler\ candidate sample investigated by \citet{fultonetal2017}, we conclude that more planets are required to quantitatively confirm the gap with \Ktwo\ candidates or validated planets. This work, in addition to increasing the population of validated \Ktwo\ planets by nearly 50\% and providing new targets for follow-up observations, will also serve as a framework for validating candidates from upcoming \Ktwo\ campaigns and the \textit{Transiting Exoplanet Survey Satellite}, expected to launch in 2018.

\end{abstract}


\section{Introduction} \label{intro}

The field of exoplanets is relatively young compared to most other disciplines of astronomy: the announcement of the first exoplanet orbiting a star similar to our own was made only in 1995\footnote{The exoplanet 51 Peg b was not fully confirmed to be a planet until the absolute mass was measured by \citet{martinsetal2015}. Moreover, a reported brown dwarf discovered in 1989 \citep{lathametal1989} may in fact be an exoplanet, depending on its inclination.} \citep{mayorandqueloz1995}. Since then, the field has expanded rapidly, with several thousand exoplanets having now been discovered. With many upcoming extremely large telescopes, the number of known exoplanets and our understanding of them will only increase.

One of the most important moments in the history of exoplanet science was the beginning of the \Kepler\ mission \citep{boruckietal2008}. Launched in 2009, the \Kepler\ space telescope observed over 100,000 stars in a single patch of sky for four years in order to look for transits. \Kepler\ has been an overwhelming success. According to the NASA Exoplanet Archive\footnote{\url{https://exoplanetarchive.ipac.caltech.edu/}} (accessed 2018 February 14), it is currently responsible for 2341 verified exoplanets, more than every other exoplanet survey combined. 

Unfortunately, in 2013 the second of four reaction wheels on the \Kepler\ spacecraft failed, preventing the spacecraft from looking at its designated field and bringing an end to the original mission. Fortunately, a follow-up mission, called K2, was developed that used the spacecraft's thrusters as a makeshift third reaction wheel \citep{howelletal2014}. Unlike the original \Kepler\ mission, the \Ktwo\ mission must observe new fields roughly every $83$ days\footnote{Roughly $75$ of those days are devoted to science.}. As a result, \Ktwo\ observations are divided into ``campaigns'' each corresponding to a field.

With every new campaign, \Ktwo\ observes more bright stars and finds more planets orbiting these stars, so there are new bright targets available for follow-up such as radial velocity measurements or transmission spectroscopy. \Ktwo\ has led to the discovery of numerous candidate and confirmed planets \citep{crossfieldetal2015, foreman-mackeyetal2015, montetetal2015, vanderburgetal2015a, vanderburgetal2016a, adamsetal2016, barrosetal2016, crossfieldetal2016, schliederetal2016, sinukoffetal2016, popeetal2016, dressingetal2017b,  hiranoetal2017, martinezetal2017}, as well as to the identification of planets orbiting rare types of stars, including particularly bright nearby dwarf stars \citep{petiguraetal2015, vanderburgetal2016b, christiansenetal2017, crossfieldetal2017, niraulaetal2017, rodriguezetal2017a, rodriguezetal2017b}, young, pre-main-sequence stars \citep{mannetal2016, davidetal2016}, and disintegrating planetary material transiting a white dwarf \citep{vanderburgetal2015b}. 

Here, we take advantage of the large number of bright stars observed by \Ktwo\ and present the identification and follow-up of a sample of \ntrescands\ exoplanet candidates orbiting stars (brighter than 13 mag) in the \Kepler\ bandpass identified from \Ktwo\ Campaigns 0 through 10. Since the beginning of the \Ktwo\ mission, we have also obtained spectra for all of our candidates, as well as many high-resolution imaging observations, in order to measure the candidate host stars' parameters and identify nearby stars (both types of follow-up aid in identification and ruling out of false-positive scenarios). We also attempt to validate\footnote{The difference between an exoplanet candidate and a validated exoplanet is very important. During the original \Kepler\ mission, an exoplanet candidate was a transit signal that had passed a battery of astrophysical false-positive and instrumental false-alarm tests. In \Ktwo, however, the usage is looser; the term is commonly used to refer to any exoplanet signal that a particular team has identified as a possible planet. So long as the reasoning is sound and the results are published, the signal is effectively a candidate. A validated planet is a candidate that has been vetted with follow-up observations and determined quantitatively to be far more likely an exoplanet than a false positive (according to some likelihood threshold). Validated planets, because confidence in their planethood is higher than for a regular candidate, are far more promising targets than planet candidates for follow-up observations, characterization, and eventual confirmation. We note that validation is not the same thing as confirmation, which is ideally attained through a reliable mass determination. In this work, we are in general not attempting to ``confirm'' planets. Confirmation is more rigorous than validation, in the same way that validation is more rigorous than candidacy. Confirmation is usually accomplished via the RV method, the TTV method, or, less commonly, methods such as a full photodynamical modeling solution \citep[e.g.][]{carteretal2011} or Doppler tomography \citep[e.g.][]{zhouetal2016}.} our candidates with \vespa, a statistical validation tool developed by \citet{morton2012, morton2015b}, finding \nplanets\ to be validated with a false-positive probability (FPP) less than $0.1\%$. Of these newly validated exoplanets, \nalreadycand\ were previously only exoplanet candidates and \nnew\ have not been previously detected. This work will increase the validated \Ktwo\ planet sample from 212 (according to the Mikulski Archive for Space Telescopes\footnote{\url{https://archive.stsci.edu/k2/published\_planets/search.php}}; accessed 2018 February 14) to 307, an increase of nearly 50\%. Similarly, this work increases the \Ktwo\ candidate sample by ${\sim}20\%$.

In this paper we describe the identification and analysis of our candidate sample, as well as our validation process and the resulting validated planet sample. Section \ref{pixelsplanets} discusses the process by which we use \Ktwo\ data to identify exoplanet candidates. Section \ref{support_obs} describes our ground-based observations of the planet candidate host stars detected by \Ktwo. Section \ref{analysis} explains the analysis of the \Ktwo\ light curves and the follow-up spectroscopy and high-contrast imaging. Section \ref{vespa} details how we use \vespa\ to calculate FPPs for our planet candidates. Section \ref{results} presents the results of the our candidate identification, vetting, follow-up observations, and analysis in detail for a single, instructive planet. Then the results for the entire planet candidate sample are similarly presented. In Section \ref{discussion}, we discuss the results of our work, including confirmation of features in the exoplanet population previously identified using data from the original \Kepler\ mission. Finally, we summarize and conclude in Section \ref{conclusion}.

\section{Pixels to planets} \label{pixelsplanets}
In this section, we first explain how \Ktwo\ observations are collected, then we describe the process by which systematic errors are removed from \Ktwo\ data, and finally we discuss the analysis of the systematics-corrected \Ktwo\ data in order to identify planet candidates.

\subsection{\Ktwo\ Observations} \label{k2_obs}

Since 2014, the \Ktwo\ mission has served as the successor to the original \Kepler\ mission. By observing fields along the ecliptic plane and firing its thrusters approximately once every six hours, the probe can maintain an unstable equilibrium against solar radiation. However, the spacecraft can only point toward a given field for roughly $83$ days before re-pointing (in order to keep sunlight on the spacecraft panels and out of its telescope).

Because of on-board data storage constraints, not all data collected by the CCD array can be retained and transmitted to the ground. As a result, targets must be identified within each campaign field prior to observation so that non-target data can be discarded and a postage stamp (a small group of pixels) around each target can be saved and transmitted to the ground.

In the original \Kepler\ mission, the primary objective was to determine the frequency of Earth-like planets orbiting Sun-like stars \citep{batalhaetal2010a}. Although some planet search targets were selected during mission adjustments and others were selected through a Guest Observer (GO) program for secondary science objectives, most targets were selected pre-launch for the primary objective. However, \Ktwo\ operates in a very different manner. For each \Ktwo\ campaign, targets are exclusively selected through the GO program, which evaluates observing proposals submitted by the astronomical community for any scientific objective, not just exoplanet-related objectives. Ideally, GO proposals have scientifically compelling goals that can be achieved through \Ktwo\ observations and cannot easily be achieved with other instruments or facilities.

In a typical \Ktwo\ campaign, the number of targets ranges between 10,000 and 40,000 with long-cadence observations (${\approx} 30$-minute integration), and about 50-200 with short-cadence observations (${\approx} 1$-minute integration). Exceptions include C0, which served primarily as a proof-of-concept campaign to show that the \Ktwo\ mission was viable, and C9, which focused on microlensing targets in the Galactic Bulge. Both C0 and C9 had fewer targets than normal in both long cadence and short cadence. It should also be noted that there are occasional overlaps between campaign fields. Despite fewer new targets, overlaps provide a longer baseline of observations for targets of interest in the overlapping region.

This paper focuses on Campaigns 0 through 10 (excluding Campaign 9). However, the process implemented in this research can easily be extended and applied to additional \Ktwo\ campaigns.

\subsection{\Ktwo\ Data Reduction} \label{k2_data_reduction}

Because of the loss of two reaction wheels, the \Kepler\ telescope is perpetually drifting off target and must be regularly corrected by thruster fires, causing shifts in the pixels that targets fall on. These shifts, coupled with variable sensitivity between pixels on the telescope CCDs and variable amounts of starlight falling inside photometric apertures, lead to systematic variations in the signal from \Ktwo\ targets, introducing noise into the photometric measurements. \citet{howelletal2014} estimated that raw \Ktwo\ precision is roughly a factor of 3-4 times worse than the original \Kepler\ precision (depending on stellar magnitude). Fortunately, an understanding of the motion of the \Kepler\ spacecraft allows for modeling and correction of the induced systematic noise. In particular, we rely on the method of systematic correction described by \citet[hereafter referred to as VJ14]{vanderburgandjohnson2014}, as well as the updates to the method described in \citet[hereafter referred to as V16]{vanderburgetal2016a}. We briefly describe here the method developed by VJ14.

\begin{figure*}[t!]
\epsscale{0.8}
  \begin{center}
      \leavevmode
\plotone{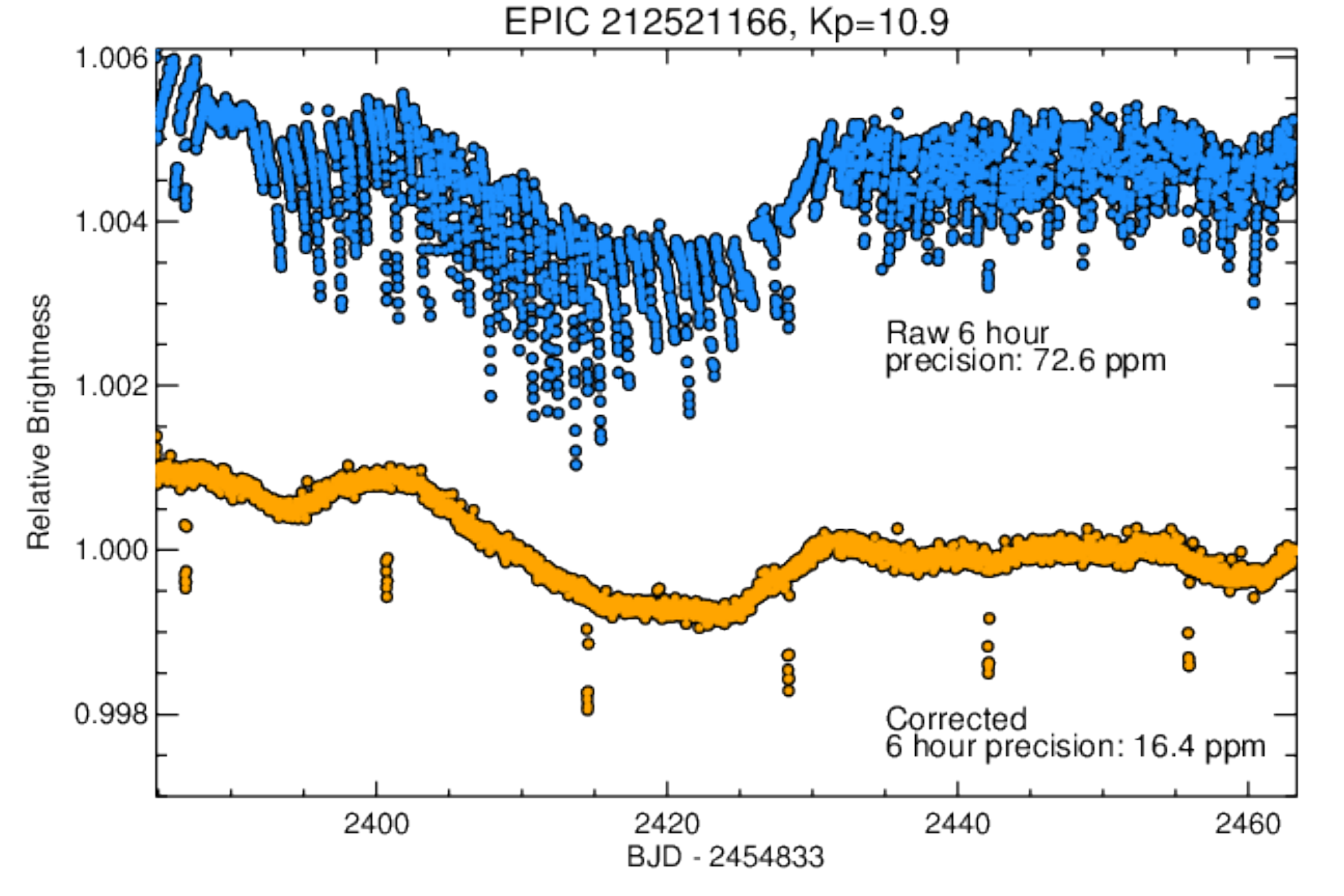}
\caption{Example of the \Ktwo\ systematics reduction process on the light curve of the planet host EPIC 212521166. The blue points show the light curve before correcting for the systematics induced by the roll motion of the \Kepler\ spacecraft, while the yellow points show the same light curve after those systematics have been removed via the data reduction process summarized in Section \ref{k2_data_reduction} and documented in \citet{vanderburgandjohnson2014} and \citet{vanderburgetal2016a}. The remaining downward dips in the corrected (yellow) light curve are transits of a mini-Neptune sized exoplanet, validated in this work and previously confirmed in \citet{osbornetal2017}.} \label{k2_lc_reduction}
\end{center}
\end{figure*}

First, 20 different aperture masks were chosen for each target star, 10 circular masks of varying size and 10 masks shaped liked the \Kepler\ pixel response function (PRF) for the target with varying sensitivity cutoffs. These masks were used to perform simple aperture photometry to produce 20 different ``raw'' light curves. Then the motion of the target star across the CCD was estimated by calculating centroid position for each cadence\footnote{Although it is possible to produce light curves by decorrelating with centroid positions measured from each star, we used the centroids measured from one hand-selected isolated, bright \Ktwo\ target per campaign, which we found gives better results.}. Next, the recurrent path of the centroid across the CCD between thruster fires was identified. Data collected during thruster fires were identified and removed. Then, for each of the 20 raw light curves produced, low-frequency variations ($> 1.5$ days) were removed with a basis spline, and the relationship between centroid position and flux was fit with a piecewise function. Because the centroid path would shift on timescales longer than 5-10 days, the flux-centroid piecewise function was applied separately to each 5-10 day light-curve segment of 5-10 days. This function was then used to correct the raw data so that low-frequency variations could be recalculated. This process was then repeated iteratively until convergence. Finally, after all 20 raw light curves per star were processed in this way, a ``best'' aperture was chosen to maximize photometric precision. An example of a light curve before and after the full data reduction procedure can be seen in Fig.~\ref{k2_lc_reduction} for the planet host EPIC 212521166. We note that light from any nearby companion stars could potentially enter the best aperture mask, which may lead to a diluted transit and an underestimated planet radius \citep{ciardietal2015, hirschetal2017}. However, we expected this effect to be small even when present and therefore do not correct for it.

\subsection{Identifying Threshold Crossing Events} \label{tces_and_triage}

AFter the roll systematics were removed from the photometry according to the method described by VJ14, we conducted a transit search of each \Ktwo\ target using the method of \citet{vanderburgetal2016a}. We give a short description of the transit search process here.

First, low-frequency variations were removed via a basis spline and outliers were removed. Then a box-least-squares (BLS) periodogram \citep{kovacsetal2002} was calculated over periods between 2.4 hr and half the length of the campaign. All periodic decreases in brightness with a signal-to-noise ratio (S/N) $> 9$ were investigated. If putative transits lasted longer than 20\% of their period, or were composed of a single data point, or changed depth by over 50\% when the lowest point was removed, the signal was removed and the BLS periodogram recalculated. Any detection passing these tests was deemed a threshold-crossing event (TCE). We identified \ntces\ TCEs across C0-C10 in this manner.

Each TCE was fit with the \citet{mandelandagol2002} transit model to estimate transit parameters, then the TCE was removed from the light curve, and the BLS periodogram was recalculated. AFter all TCEs had been identified, they subsequently underwent ``triage'', in which each candidate was inspected by eye in order to remove obvious astrophysical false positives and instrumental false alarms from subsequent analysis. TCEs identified as neither type of false signal passed the triage phase and moved on to a ``vetting'' phase. 

\subsection{Identifying \Ktwo\ Candidates} \label{candidates_and_vetting}

\begin{figure*}[t!]
\epsscale{1.0}
  \begin{center}
      \leavevmode
\plotone{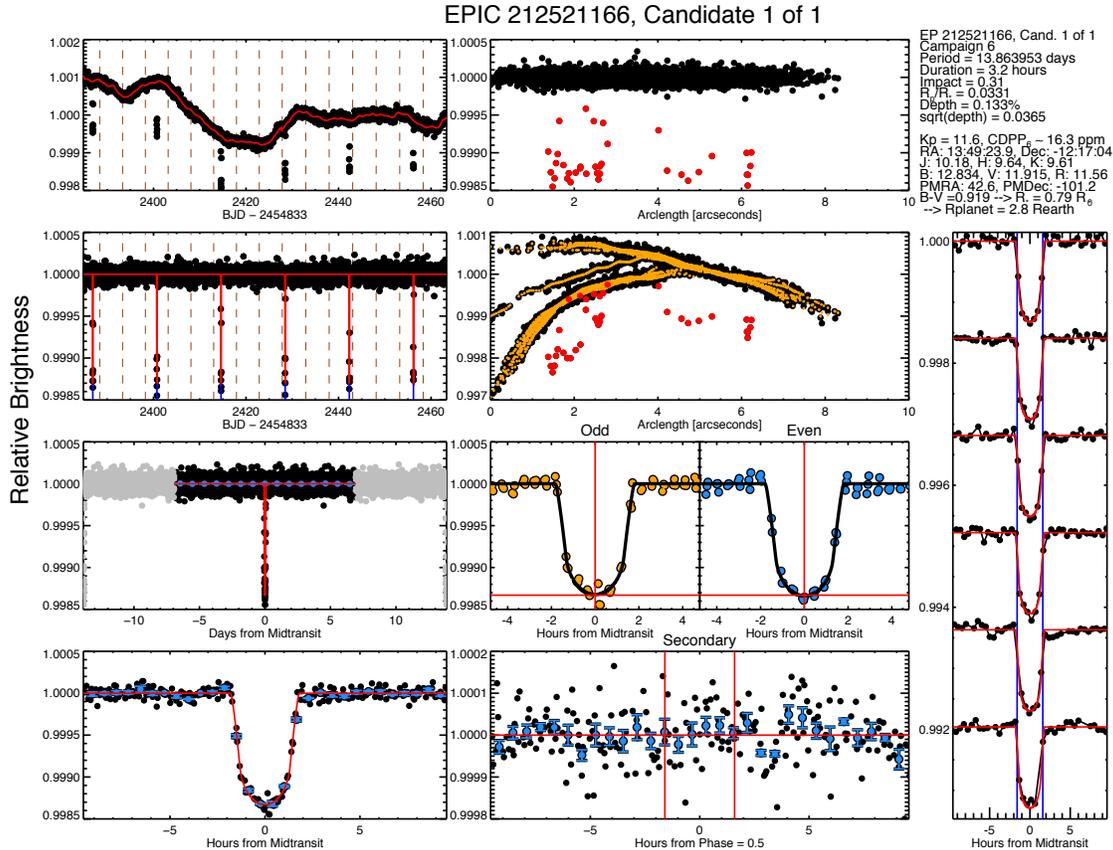}
\caption{Diagnostic plots for EPIC 212521166.01. Left column, first (top) and second rows: \Ktwo\ light curves without and with low-frequency variations removed, respectively. The low-frequency variations alone are modeled in red in the first row, whereas the best-fit transit model is shown in red in the second row. Vertical brown dotted lines denote the regions into which the light curve was separated to correct roll systematics. Left column, third and fourth rows: phase-folded, low-frequency corrected \Ktwo\ light curves. In the third row, the full light curve is shown (points more than one half-period from the transit are gray), whereas in the fourth row, only the light curve near transit is shown. The red line is the best-fit model, and the blue points are binned data points. Middle column, first and second rows: arclength of centroid position of star versus brightness, after and before roll systematics correction, respectively. Red points denote in-transit data. In the second row, small orange points denote the roll systematics correction made to the data. Middle column, third row: separate plotting and modeling of odd (left panel) and even (right panel) transits, with orange and blue data points, respectively. The black line is the best-fit model, the horizontal red line denotes the modeled transit depth, and the vertical red line denotes the mid-transit time (this is useful for detecting binary stars with primary and secondary eclipses). Middle column, fourth row: light curve data in and around the expected secondary eclipse time (for zero eccentricity). Blue data points are binned data, the horizontal red line denotes a relative flux = 1, and the two vertical red lines denote the expected beginning and end of the secondary eclipse. Right column: individual transits (vertically shifted from one another) with the best-fit model in red and the vertical blue lines denoting the beginning and end of transit.} \label{vetting1}
\end{center}
\end{figure*}

\begin{figure*}[t!]
\epsscale{1.0}
  \begin{center}
      \leavevmode
\plotone{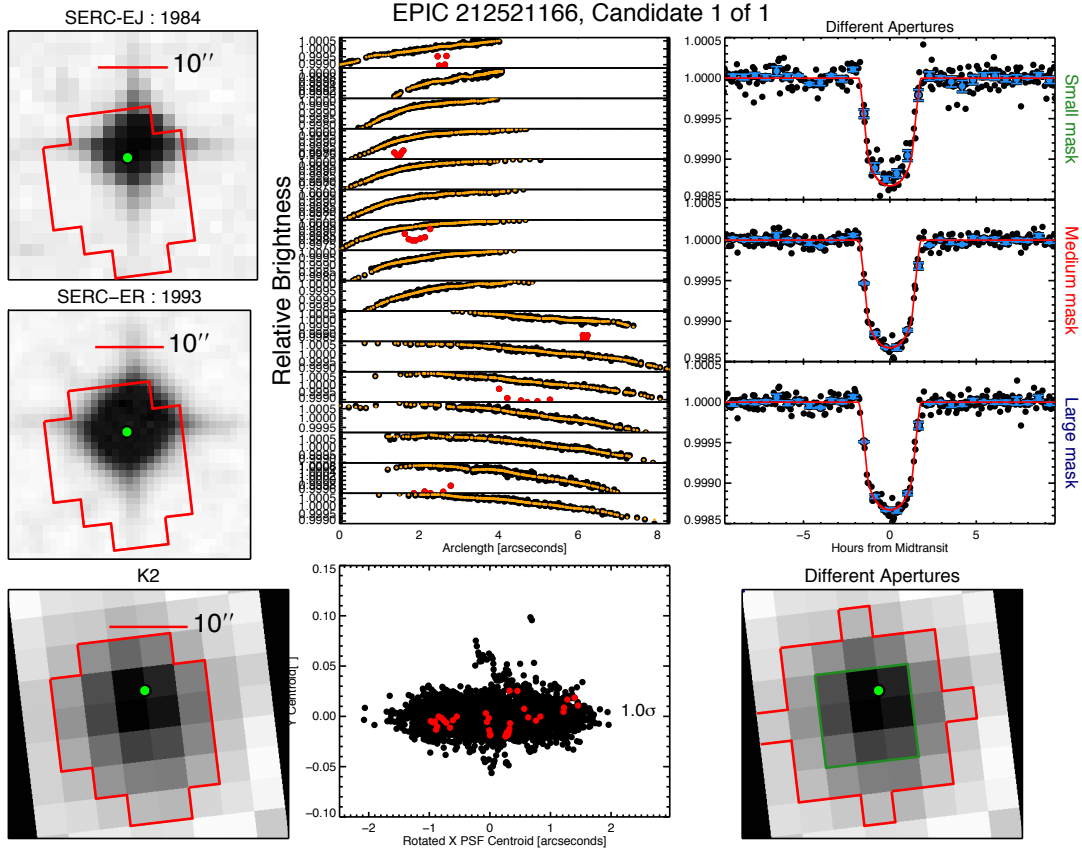}
\caption{Diagnostic plots for EPIC 212521166.01. Left column, first (top), second, and third rows: images from the first Digital Sky Survey, the second Digital Sky Survey, and \Ktwo\ respectively, each with a scale bar near the top and an identical red polygon to show the shape of the photometric aperture chosen for reduction. The \Ktwo\ image is rotated into the same orientation as the two archival images (north is up). Middle column, top row: multiple panels of uncorrected brightness versus arclength, chronologically ordered and separated into the divisions in which the roll systematics correction was calculated. In-transit data points are shown in red, orange points denote the brightness correction applied to remove systematics. Middle column, bottom row: variations in the centroid position of the \Ktwo\ image. In-transit points are red. The discrepancy (in standard deviations) between the mean centroid position in transit and out-of-transit is shown on the right side of the plot. Right column, first row: the \Ktwo\ light curve near transit as calculated using three differently sized apertures: small mask (top panel), medium mask (middle panel), and large mask (bottom panel), each with the identical best-fit model in red. Aperture-size dependent discrepancies in depth could suggest background contamination from another star. Right column, third row: the \Ktwo\ image overlaid with the three masks from the previous plot shown (in this figure, the large mask is fully outside the postage stamp and is therefore not visible).} \label{vetting2}
\end{center}
\end{figure*}

\begin{figure*}[t!]
\epsscale{0.95}
  \begin{center}
      \leavevmode
\plotone{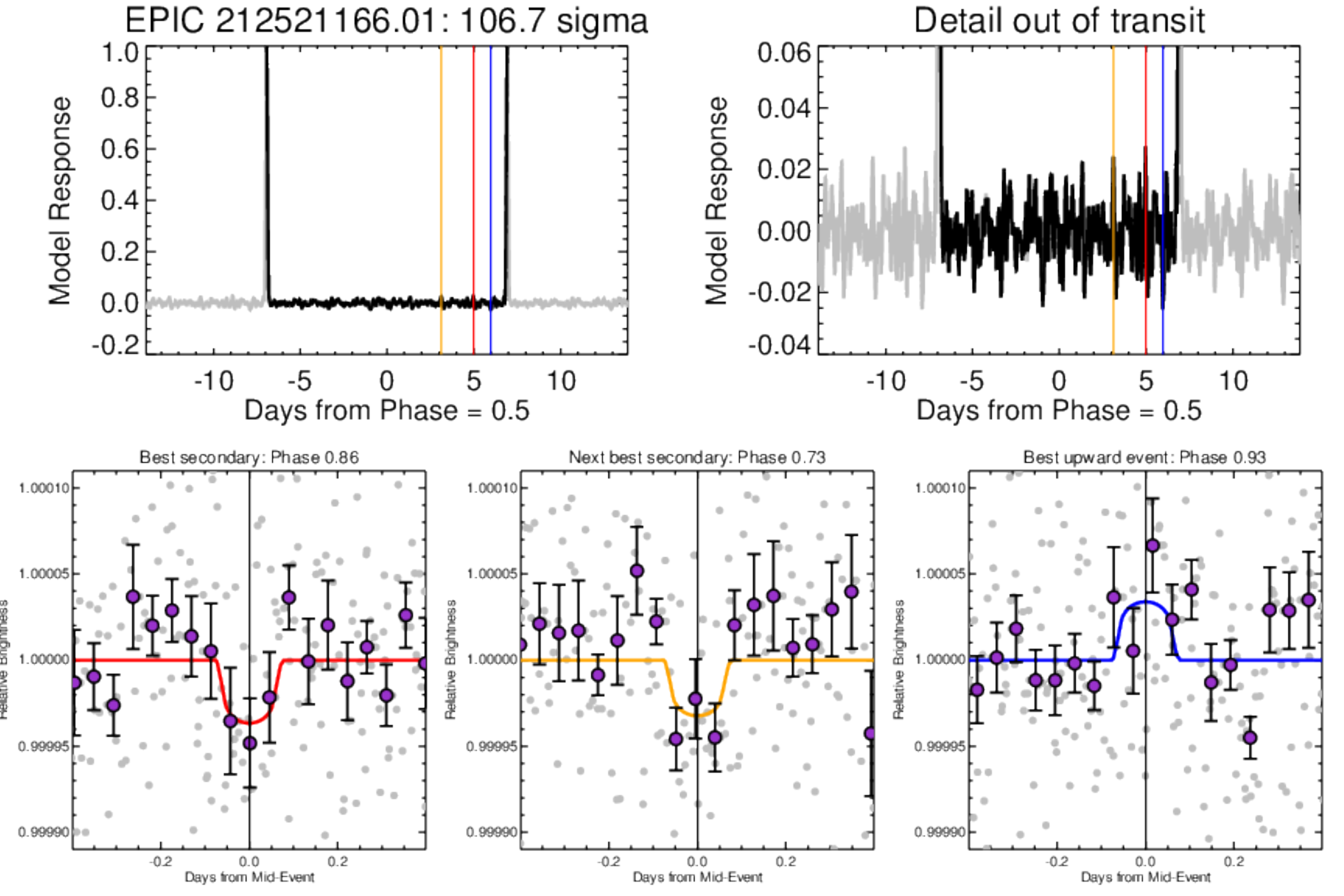}
\caption{Model-shift uniqueness diagnostic plots for EPIC 212521166.01. We cross-correlated the phase-folded light curve of each planet candidate with the candidate's best-fitting transit model for the candidate. The two plots in the top row show the cross-correlation function as a function of orbital phase with different scaling on the Y-axis. The tallest peak in the response is the primary transit, and the three vertical lines show the phase of the highest significance putative secondary event (red) , the second-highest significance putative secondary event (orange), and the highest significance upside-down secondary event (blue). The bottom row of plots show the three segments of the light curve surrounding the highest and second-highest significance putative secondary eclipses, and the highest significance upside-down event. Gray and purple data points are unbinned and binned observations, respectively. Comparing the amplitudes of these three events gives an indication of their significance. Here, the two most significant putative secondary eclipses have a similar amplitude to the most significant upside-down event, indicating that they are all likely spurious. We find no significant secondary eclipse in the light curve of EPIC 212521166.01.} \label{modelshift}
\end{center}
\end{figure*}

During vetting, we subjected the surviving TCEs to a battery of additional tests to identify astrophysical false positives and instrumental false alarms. Some of these tests were identical or similar to the tests conducted for the \Kepler\ mission, while others are specific to \Ktwo\ data. For each test we produced a diagnostic plot, examples of which are shown in Figs.~\ref{vetting1}, \ref{vetting2}, and \ref{modelshift}. Here we describe the tests we conducted in more detail.

\begin{enumerate}
    \item The times of the in-transit points of a TCE were compared against the position of the \Kepler\ spacecraft at these times, as many instrumental false alarms were composed of data points near the edges of \Kepler's rolls where the \Ktwo\ flat field is less well constrained, and our analysis method can leave in systematics. The plots we used to identify these false alarms are shown in Figs.~\ref{vetting1} and \ref{vetting2}.
    \item We compared the signal of a TCE in light curves produced using multiple different photometric apertures. This test is a powerful way to identify signals caused by instrumental systematics (as these systematics present differently in different photometric apertures), as well as identifying astrophysical false positives, such as when a candidate transit signal was due to contamination from a nearby star. An example of these tests is shown in Fig.~\ref{vetting2}. We note that although this test rules out transits or eclipses originating from a nearby companion, it does not rule out the possibility of light contamination from a nearby companion, which could dilute the observed transit and lead to an underestimation of the planetary radius \citep{ciardietal2015, hirschetal2017}.
    \item Individual transits of a TCE were visually inspected, since instrumental false alarms were less likely to have consistent, planet-like transit depths or shapes (see Fig.~\ref{vetting1}). This metric is qualitatively similar to the ``transit patrol'' metrics introduced in the DR25 \Kepler\ planet candidate catalog \citep[in particular the Rubble, Marshall, Chases, and Zuma tests;][]{thompsonetal2017}.
    \item Flux-centroid motion, phase variations (possibly caused by relativistic beaming, reflected light, or ellipsoidal effects), differences in depth between odd- and even-numbered transits, and secondary eclipses were all searched for as evidence of astrophysical false positives. Similar tests have been used since the beginning of the \Kepler\ mission \citep{batalhaetal2010b}. Example diagnostic plots for identifying phase variations, the difference between even- and odd-numbered transits, and secondary eclipses near phase 0.5 are shown in Fig.~\ref{vetting1}; flux-centroid shift tests are shown in Fig.~\ref{vetting2}. We searched for secondary eclipses at arbitrary phases (not necessarily near phase = 0.5) using the model-shift uniqueness test designed by \citet{coughlinetal2016}. We show diagnostic plots for the model-shift uniqueness test in Fig.~\ref{modelshift}.
    \item We searched for astrophysical false-positive scenarios by ephemeris matching. Sometimes the pixels surrounding a target star can be contaminated with a small amount of light from other nearby stars. When those nearby stars are variable themselves (like eclipsing binaries), the variability from the nearby stars can be introduced into the target's light curve \citep{coughlinetal2014}. We identified cases where this happened by searching for planet candidates that have the same period (or integer multiples of the same period) and time of transit as eclipsing binaries observed by \Ktwo\ using the same criteria as \citet{coughlinetal2014}. We found no examples of matched ephemerides due to direct PRF overlap that we had not also identified in our tests with multiple photometric apertures, but we did find two instances where candidate transit signals were caused by charge transfer inefficiency along one of the columns of the \Kepler\ detector. We excluded a candidate around EPIC 212435047 caused by contamination from the eclipsing binary EPIC 212409377 located along the same CCD column about 2000 arcsec away, and we excluded a candidate around EPIC 202710713 contaminated by the eclipsing binary EPIC 202685801 located along the same CCD column at a distance of about 300 arcsec.
    \item We estimated the S/N of a TCE by taking the difference between the mean baseline flux and the in-transit flux and dividing by the quadrature sum of the standard deviation of the flux in those two regions. We defined the boundaries of the in-transit region in two ways and calculated the S/N for both choices. In one case we used every data point collected between the second and third contact (minus a single long-cadence Kepler exposure on either side). This estimate of S/N could sometimes not be calculated if the transit was grazing. In the other case we used the central 20\% of data collected from the first to fourth contact (which could be calculated even if the transit was grazing). Following standard practice from the Kepler mission, if both of these values (or just the latter in the grazing case) were below $7.1 \sigma$ the candidate was excluded \citep[$7.1 \sigma$ is the minimum significance level for a signal to qualify as a TCE in the \Kepler\ Science Pipeline;][]{jenkinsetal2010}. We only encountered two such candidates: EPIC 220474074 and EPIC 201289302.

\end{enumerate}

Although it is possible to automate diagnostic tests of this type \citep[see, for example, the Robovetter that was designed for the main \Kepler\ mission;][]{coughlinetal2016}, we performed vetting tests 1-4 by eye.

Any TCE surviving all of these vetting stages was promoted to planet candidate (\ncands\ were promoted in this way). All candidates orbiting sufficiently bright host stars (see Section \ref{TRES}) were then subjected to our validation process.

\section{Supporting Observations} \label{support_obs}

In this section, we describe the follow-up observations we conducted to better characterize the candidate host stars. These observations are crucial for improving stellar parameters \citep[e.g.][]{dressingetal2017a, mannetal2017, martinezetal2017}, which in turn can help differentiate between a transiting planet and various false-positive scenarios (i.e. stellar binary configurations) that might prefer different regions of stellar parameter space. We first discuss high-resolution optical spectroscopy of the planet candidate host stars from the Tillinghast Reflector Echelle Spectrograph (TRES), followed by speckle imaging from the Differential Speckle Survey Instrument (DSSI) and the NASA Exoplanet Star and Speckle Imager (NESSI) at the WIYN telescope,\footnote{The WIYN Observatory is a joint facility of the University of Wisconsin-Madison, Indiana University, the National Optical Astronomy Observatory, and the University of Missouri.} the Gemini South telescope, and the Gemini North telescope, and finally, adaptive optics (AO) imaging from Keck Observatory, Palomar Observatory, Gemini South Observatory, Gemini North Observatory, and the Large Binocular Telescope Observatory.

\subsection{TRES Observations} \label{TRES}

All of the spectra used in this work were obtained with TRES, a spectrograph with a resolving power of $R = 44,000$ and one of two spectrographs for the 1.5 m Tillinghast telescope at the Whipple Observatory on Mt. Hopkins in Arizona. We obtained at least one usable TRES spectrum of each of the planet candidate host stars that we consider in this work and that we subject to our validation process (see Section \ref{stellar_parameters} for our definition of ``usable''). With a few exceptions, we only observed candidates brighter than 13 mag in the \Kepler\ band with TRES because of the lengthy integration time required to collect spectra of stars fainter than this and the difficulty of subsequent follow-up observations (for example, with precise radial velocities) at other facilities. This limitation reduced the number of candidates we considered for validation significantly, from \ncands\ to \ntrescands. In the future, observing these fainter candidates, either with TRES or other spectrographs on larger telescopes, could potentially more than double the number of \Ktwo\ planets for our analysis.

% %\begin{comment}
% %\tabletypesize{\footnotesize} 
% %\fontsize{9}{11}\selectfont
% \LongTables
% %\begin{table}[h] 
% %\centering
% \begin{deluxetable}{ccc}
% %\begin{table}
% %\tablecolumns{8} 
% %\tablewidth{0pt}
% \tablecaption{High-resolution Imaging
% \label{table:ao_speckle_table}} 
% \tablehead{ 
% EPIC & Filter & Instrument}
% \startdata 
% 201110617 & 562 & NESSI \\
% 201110617 & 832 & NESSI \\
% 201111557 & 562 & NESSI
% \enddata
% %end{table}
% \tablenotetext{}{(This table is available in its entirety in machine-readable form.)}
% \end{deluxetable}
% %\end{comment}

\subsection{Speckle Observations} \label{speckle}

We observed many of our planet candidates with speckle imaging from either the 3.5 m WIYN telescope, the Gemini-South 8.1 m telescope, or the Gemini-North 8.1 m telescope. Together, the three telescopes collected 162 speckle images of 73 stars with DSSI \citep{horchetal2009}. DSSI is a speckle-imaging instrument that travels between different telescopes. For each of the 73 targets, we collected DSSI speckle images at narrowband filters centered at 6920 and 8800 \AA\ (at least one of each for every target). These observations were made in 2015 September and October, as well as in 2016 January, April, and June.

Furthermore, 160 speckle images were collected for a distinct sample of 70 stars at the WIYN telescope using NESSI, which is essentially a newer version of the DSSI instrument. For each of the 70 targets, we collected NESSI speckle images at narrowband filters centered at 5620 and 8320 \AA\ (at least one of each). These observations were made in 2016 October through November and 2017 March through May. A list of the observed stars can be found in Table~\ref{table:ao_speckle_table}.

\subsection{AO Observations} \label{ao}

In addition to speckle imaging, we also observed many of our planet candidate host stars with AO imaging.

We collected 47 AO images for 45 stars on the Keck II 10 m telescope in K filter with the Near Infra Red Camera 2 (NIRC2); 5 of these stars were also imaged using NIRC2 in $J$ band. All of these observations were made during 2015 April, July, August, and October as well as 2016 January and February.

We collected 27 AO images for 27 stars on the Palomar 5.1 m Hale telescope in K filter with the Palomar High Angular Resolution Observer \citep[PHARO,][]{hawyardetal2001}; 6 of these stars were also imaged using PHARO in $J$ band. All of these observations were made during 2015 February, May, and August as well as 2016 June, September, and October.

We collected 19 AO images for 18 stars on the Gemini-North 8.1 m telescope in $K$ band with the Near InfraRed Imager and spectrograph \citep[NIRI,][]{hodappetal2003}. These observations were made during 2015 October and November as well as 2016 June and October.

We collected a single AO image on the Large Binocular Telescope in K filter with the L/M-band mid-infrared Camera \citep[LMIRCam,][]{leisenringetal2012}. This observation was made in 2015 January.

There was some overlap between instruments; overall, AO images were collected for a total of 80 systems. A list of the observed stars can be found in Table~\ref{table:ao_speckle_table}.

\section{Data Analysis} \label{analysis}

After all of the photometry had been reduced and all of the necessary follow-up observations had been collected, the next step was to analyze the data, calculate relevant parameters, and prepare the results for the validation process. In this section, we explain the process of fitting a model to our reduced light curves (to determine transit parameters and create folded light curves), analyzing our spectra (to calculate stellar parameters), and extracting and reducing data from our high-contrast images (to create contrast curves).

\subsection{\Ktwo\ Light Curves} \label{k2_light_curves}

\subsubsection{Simultaneous Fitting of \Ktwo\ Systematics and Transit Parameters} \label{simultfit}

In Sections \ref{k2_data_reduction}, \ref{tces_and_triage}, and \ref{candidates_and_vetting} we described the process of correcting \Ktwo\ photometry for instrumental systematics and exploring the reduced light curves for candidates. After these steps were completed, the planet candidates needed to be more thoroughly characterized. In order to assess transit and orbital parameters, we reproduced the \Ktwo\ light curves for these planet candidates by rederiving the systematics correction while simultaneously modeling the transits in the light curve. As in our original systematics correction, the light curve was divided into multiple sections and the systematics correction was applied to each section separately. A piecewise linear function was fit with breaks roughly every $0.25$ arcsec (varying slightly by target) to the arclength versus brightness relationship described in section \ref{k2_data_reduction} (arclength is a one-dimensional measure of position along the path an image centroid traces out on the \Kepler\ CCD camera). The low-frequency variations in the light curve were modeled with a cubic spline (with breakpoints every 0.75 days), and the transits themselves were modeled with the \citet{mandelandagol2002} transit model. The fit was performed using a Levenberg-Marquardt optimization \citep{markwardt2009}, and all of the parameters from the optimization (besides the transit parameters) were used in order to correct the systematics of the light curve (once again) and remove the low-frequency variations (once again). Since these parameters were determined in a \textit{simultaneous} fit with the transits, the quality of the resulting light-curve reduction tended to be better than that of the original light curves. 

\subsubsection{Final Estimation of Transit Parameters and Uncertainties} \label{transit_model}

After we produced the systematics-corrected, low-frequency-extracted light curves, we analyzed them further in order to estimate final transit parameter values and their uncertainties. We based our model on the \texttt{BATMAN} Python package \citep{kreidberg2015}, which we used to calculate our synthetic transit light curves. We fit the transit light curves of all planet candidates around a given star simultaneously so that overlapping transits could be modeled, assuming that each of the planets were non-interacting and on circular orbits. For each planet candidate, five parameters were included: the epoch (i.e. time of first transit), the period, the inclination, the ratio of planetary to stellar radius ($R_\mathrm{p}/R_*$), and the semi-major axis normalized to the stellar radius ($a/R_*$). Additionally, two parameters for a quadratic limb-darkening law were included \citep{kipping2013}, as well as a parameter to allow the baseline to vary (in case there was an erroneous systematic offset from flux = 1 outside of transit), and a noise parameter that assigned the same uncertainties to each flux measurement (since flux error bars were not calculated in the \Ktwo\ data reduction process). For all of these planet and system parameters we assumed a uniform prior, except for the $R_\mathrm{p}/R_*$ parameter for each planet, which we gave a log-uniform prior.

\begin{figure}[t!]
\epsscale{1.2}
  \begin{center}
      \leavevmode
\plotone{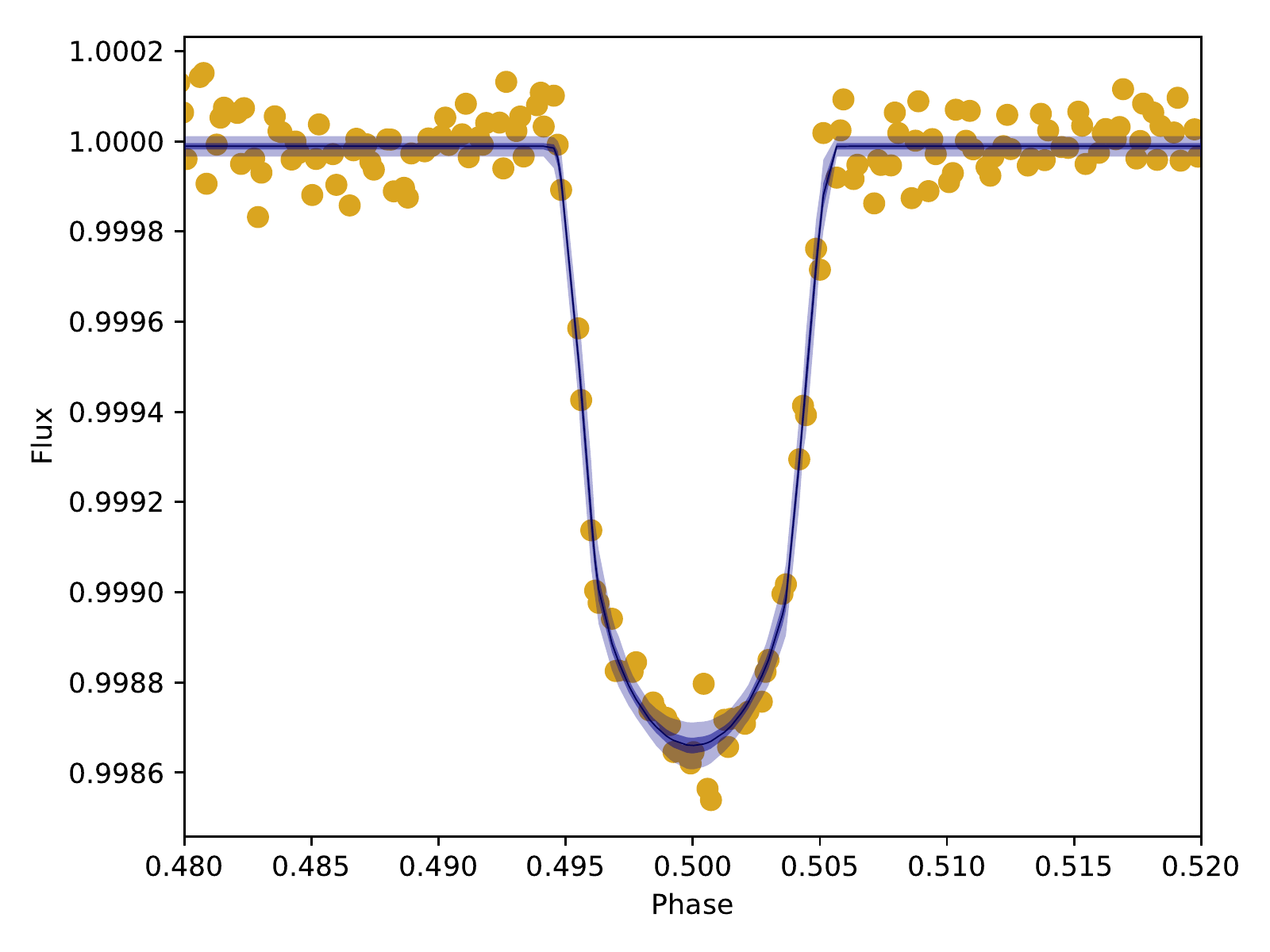}
\caption{Fit of our transit model to corrected and normalized light curve data for EPIC 212521166.01. The yellow points are the observed data, normalized and phase-folded to the orbital period of the planet. The black line is our transit model with the median parameter values determined from an MCMC process. The dark blue and light blue regions are $1\sigma$ and $3\sigma$ confidence intervals, respectively.} \label{EPIC212521166_transit}
\end{center}
\end{figure}

For each candidate system, the transit parameters in this model were estimated using \texttt{emcee} \citep{foreman-mackeyetal2013}, a Python package that runs simulations using a Markov chain Monte Carlo (MCMC) algorithm with an affine-invariant ensemble sampler \citep{goodmanandweare2010}. In each simulation, the parameter space for a system with $n$ candidates was sampled with $8+10n$ chains (equal to twice the number of model parameters). The MCMC process was run for either 10,000 steps or until convergence, whichever came last. Convergence was defined according to the scale-reduction factor \citep{gelmanandrubin1992}, a diagnostic that compares variance for individual chains against variance of the whole ensemble. A simulation was considered converged when the scale-reduction factor was less than 1.1 for each parameter. The Gelman-Rubin scale-reduction factor is properly defined for chains from distinct, non-communicating MCMC processes; however, we found that our simulations visually converged many times more quickly than the Gelman-Rubin diagnostic, so we decided the diagnostic would be sufficient for our purposes. An example of a converged model fit against transit data can be seen in Fig.~\ref{EPIC212521166_transit}.

Additionally, each simulation was checked after the minimum number of steps (10,000) and at the end of the simulation for any chains in the ensemble that could be easily categorized as ``bad,'' i.e. trapped in a minimum of parameter space with a poorer best fit than the minimum of the ensemble majority. In detail, a chain was classified as ``bad'' if both of the following applied:
\begin{enumerate}
    \item The Kolmogorov-Smirnov statistic between the chain with the highest likelihood step and the chain in question was greater than 0.5.
    \item $\theta_\mathrm{l} / \theta_\mathrm{g} < 1/(100 n)$, where $\theta_\mathrm{l}$ is the local maximum likelihood (the maximum likelihood within the chain in question), $\theta_\mathrm{g}$ is the global maximum likelihood throughout the ensemble, and n is the total number of steps. This test is our own invention, which is more likely to classify a chain as bad when its maximum likelihood is farther from the global maximum likelihood. This test also accounts for the length of the simulation, since the maximum likelihood within each chain should be closer to the global maximum likelihood for longer simulations. We include a factor of 100 in the denominator to make the test sufficiently conservative, such that it only finds bad chains very rarely and only those extremely unlikely to rejoin the ensemble in the lifetime of the simulation.
\end{enumerate}

If a chain was deemed bad after 10,000 steps, its position in parameter space was updated to that of a good chain from the previous step of the simulation. If a chain was deemed bad at the end of the simulation, it was simply removed and not replaced. Both after 10,000 steps and at the end of the simulation, bad chains only occurred about 15\% of the time, typically for only one or two chains in the ensemble.

A representative sample of the converged posteriors for each exoplanet system transit fit has been archived at \url{\texttt{https://zenodo.org/record/1164791}}.

\subsection{TRES Spectroscopy} \label{stellar_parameters}

We first visually inspected each of the spectra collected with TRES for our candidate host stars. Our main goal in visually inspecting the spectra was to identify cases in which the spectra were composite or otherwise indicative of multiple stars in the system. We looked at diagnostic plots produced by the TRES pipeline, which showed various echelle orders and a cross-correlation between a synthetic template spectrum and a single spectral order around 5280 \AA.

After identifying composite spectra and confirming that the others were apparently single-lined, we prepared the spectra before determining the host stars' parameters. In particular, we manually removed cosmic rays from the spectra that might bias or otherwise affect the spectroscopic parameters we measured. We focused on three echelle orders in particular, orders 22, 23, and 24 (covering 5059-5158, 5135-5236, and 5214-5317 \AA, respectively), which are within the range we used to measure spectroscopic parameters. 

After we visually inspected each spectrum and removing cosmic rays, we analyzed each TRES spectrum using the Stellar Parameter Classification (SPC) tool, developed by \citet{buchhaveetal2012}. SPC determines key stellar parameters through a comparison of the input spectrum (between 5050 and 5360 \AA) to a library grid of synthetic model spectra, developed by \citet{kurucz1992}. The library is 4-dimensional, varying in effective temperature $T_{\mathrm{eff}}$, surface gravity $\log{g}$, metallicity [m/H], and line broadening (a good proxy for projected rotational velocity, or $v\sin{i}$). [m/H] is estimated rather than [$\mathrm{Fe}/\mathrm{H}$] because \textit{all} metallic absorption features are used to determine metallicity rather than just iron lines. (SPC assumes that all relative metal abundances are the same as in the Sun, and [m/H] simply scales all solar abundances by the same factor.) This library grid spans $T_{\mathrm{eff}}$ from 3500 K to 9750 K in 250 K increments, $\log{g}$ from 0.0 to 5.0 (cgs) in 0.5 increments, [m/H] from -2.5 to +0.5 in 0.5 dex increments, and line broadening from 0 km s$^{-1}$ to 200 km s$^{-1}$ in progressively spaced increments (from 1 km s$^{-1}$ up to 20 km s$^{-1}$). In total, the library contains 51,359 synthetic spectra. The stellar parameters derived using SPC can be found in Table~\ref{table:stellar_parameters_table}.

We also calibrated TRES against \citet{duncanetal1991} for the $S_{HK}$ stellar activity index, a measure of the strength of emission in the cores of the H and K Ca II spectral absorption features. Our calibration was only performed over a range of $0.244 < $ B-V $ < 1.629$, $0.055 < $ $S_{HK}$ $ < 2.070$, and $v\sin{i}$ $<$ 20 km s$^{-1}$. Additionally, we required a photon count of more than 250,000 in the $R$ and $V$ continuum regions ($3901.07 \pm 10$ \AA\ and $4001.07 \pm 10$ \AA, respectively). Within these ranges, we were able to calculate $S_{HK}$ for 28 of our spectra collected for four stars. Although there are relatively few \Ktwo\ planet candidates bright enough for this measurement on a 1 m telescope, it will be much more common with the \textit{Terrestrial Exoplanet Survey Satellite} (TESS). A note has been made for the relevant stars in Table~\ref{table:stellar_parameters_table} and the full results are listed in Table~\ref{table:shk_table}. There is also a detailed description of the $S_{HK}$ calibration process for TRES in the Appendix.

In cases where we observed a candidate host star more than once with TRES, we attempted to identify or rule out large velocity variations indicating a stellar companion to the host star. (For this purpose we collected one new TRES spectrum each for EPIC 220250254 and EPIC 210965800, but otherwise exclusively used the spectra from which we derived stellar parameters.) We phased the radial velocities to the photometric ephemeris and fit a sinusoid to the velocities at the period and ephemeris of each candidate (and assuming a circular orbit) in order to estimate the companion mass and mass uncertainty. If the companion mass was more than $3\sigma$ lower than 13 Jupiter masses, we ruled out the eclipsing binary scenario (see Section~\ref{vespa_application}). If the semi-amplitude of the companion mass was more than 1 km s$^{-1}$ we labeled the system as a binary. Five systems were labeled binaries in this manner. If neither of the above cases was true, no action was taken.

Out of the \ntrescandsys\ candidate host stars, 43 had more than one TRES spectrum available. The stellar parameters calculated for the spectra of such candidates were combined via a weighted average based on the peak height of the cross-correlation function. The scatter between multiple spectra for the same candidate was not large relative to the assumed systematic errors. For example, \citet{buchhaveetal2012} set an internal error floor of 50 K for effective temperature; we found an average standard deviation of 40 K between spectra of the same candidate, and the scatter was less 100 K for ${\sim}90\%$ of candidates with multiple spectra.

There were some instances in which a TRES spectrum was not considered good enough for stellar parameter estimation with SPC. In particular, we did not use any spectrum for which the S/N per resolution element was $< 20$ or the cross-correlation function peak height of the spectrum was $< 0.8$. Furthermore, we also did not use any spectra that yielded SPC results outside of certain trustworthy ranges. Specifically, we only used spectra that yielded $4250 < T_{\rm eff} < 6500$ and line broadening $< 50$ km s$^{-1}$. It should also be noted that all spectra used to determine stellar parameters were collected with TRES on or before 2017 June 10 (although some usable spectra have since been collected, they were not retroactively included in our stellar parameter analysis). No planet candidate underwent the validation process unless their host star had at least one spectrum (collected with TRES) that satisfied these criteria.

\subsection{Contrast Curves} \label{contrast_curves}

\begin{figure*}[h!]
\epsscale{0.9}
  \begin{center}
      \leavevmode
\plotone{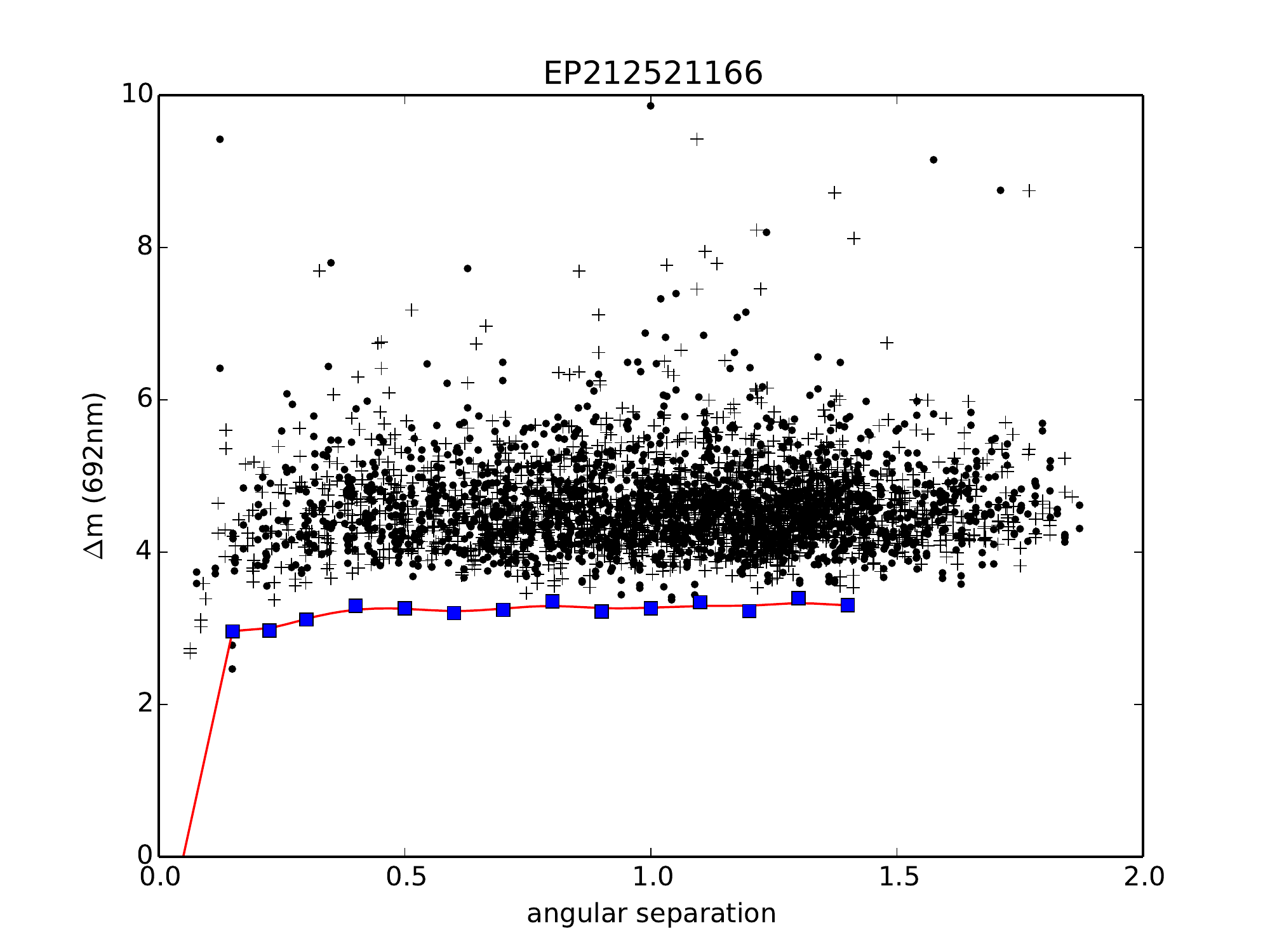}
\caption{Contrast curve developed from a speckle image at 692 nm (6920 \AA) for candidate system EPIC 212521166. The contrast curve is plotted as $\Delta m$ (average sensitivity to a companion in magnitude difference between the primary and secondary) versus angular separation (in arcseconds). Plus signs and circles are maxima and minima, respectively, of the background sensitivity in the image. The blue squares are the $5\sigma$ sensitivity limit within adjacent angular separation annuli. The red line is the contrast curve itself, a spline fit to the sensitivity limit (the blue squares).} \label{EPIC212521166_cc}
\end{center}
\end{figure*}

We quantified the constraints placed on the presence of additional companions in our high-resolution (speckle and AO) imaging by calculating contrast curves. A contrast curve specifies at a given distance from the target star how bright a companion star would have to be in order to be detected.

After the speckle observations were collected, the data were reduced according to the method described in \citet{furlanetal2017}, yielding a high-contrast image of a candidate. Then local minima and maxima were analyzed relative to the star's peak brightness to determine $\Delta m$ (average sensitivity) and its uncertainties within bins of radial distance from the target star. The contrast curves used in this research were the $5\sigma$ upper limit on $\Delta m$ as a function of radial distance. A more detailed description of the data reduction process can be found in \citet{howelletal2011}. An example of a contrast curve can be seen in Fig.~\ref{EPIC212521166_cc}.

As for AO observations, the creation of corresponding contrast curves was performed by injecting fake sources into the images and varying their brightness in order to determine the brightness threshold for the detection algorithm being used. This process is described in greater detail in \citet{crossfieldetal2016} and \citet{ziegleretal2017}

Note that these processes of creating a contrast curve were the same regardless of wavelength. So in the instances that multiple speckle or AO images at different wavelengths were available, multiple contrast curves were created that were all used in the validation process for a given candidate.

\section{False-positive Probability (FPP) Analysis} \label{vespa}

\subsection{The Application of \vespa\ to Planet Candidates} \label{vespa_application}

Our validation work relied primarily on a method called validation of exoplanet signals using a probabilistic algorithm, or \vespa, a public package \citep{morton2015b} based on the work of \citet{morton2012}. It operates within a Bayesian framework and calculates the FPP, the probability that a candidate is an astrophysical false positive rather than a true positive (i.e. a planet).

After photometry, spectroscopy, and any available high-contrast imaging had been collected and reduced for a candidate system, that system underwent our validation procedure using \vespa. For each candidate, the following information was supplied:
\begin{enumerate}
    \item A folded light curve containing the planetary transit and roughly one transit duration of baseline on either side. Identical error bars were assigned to all data points based on the noise parameter determined by our transit-fitting procedure. From this light curve, other planets in the system were removed using the best-fit parameters determined by the fitting procedure.
    \item A secondary threshold limit. We calculated this limit by first cutting out all transits from all candidates in a system. Then, for a given candidate, we phase-folded the data to the candidate's period, binned the baseline flux according to the transit duration of the candidate, calculated the standard deviation between the bin averages, and multiplied by three. Effectively, we assert that a secondary eclipse has not been detected at a $3\sigma$ level or higher.
    \item A contrast curve. When available, this lowered the FPP by eliminating the possibility of bound or background stars above a certain brightness at a given projected distance, thus reducing the parameter space in which a false-positive scenario could exist. Note that \vespa\ does not \textit{require} a contrast curve to operate, and in many cases, a candidate host star had no contrast curve data available. However, contrast curves were collected from AO or speckle data and provided to \vespa\ for \nspeckleaosys\ of our \ntrescandsys\ candidate host stars (see Table~\ref{table:ao_speckle_table}).
    \item R.A. and Decl. The likelihood of a false-positive scenario is calculated by \vespa\ based on the target star's location on the celestial sphere. For example, near the Galactic plane, a target star's FPP will increase significantly due to the crowded field.
    \item $T_{\mathrm{eff}}$, $\log{g}$, and [m/H] (collected from SPC). These constraints help rule out false-positive scenarios that would otherwise be allowed given only stellar magnitude information. Although \vespa\ \textit{could} operate without these values, we did not apply \vespa\ to any candidate system for which stellar parameters could not be estimated from a TRES spectrum.
    \item Stellar magnitudes. The \Ktwo\ Ecliptic Plane Input Catalog \citep{huberetal2016} was queried to find magnitude information on each target star in the \Kepler, $J$, $H$, and $K$ bandpasses. The $J$, $H$, and $K$ bandpasses originated from the 2MASS catalog \citep{cutrietal2003, skrutskieetal2006}. Magnitude uncertainties were not required or provided for the Kepler bandpass, while uncertainties for the $J$, $H$, and $K$ bandpasses were added in quadrature with 0.02 mag of systematic uncertainty to be conservative.
    \item Parallaxes. When parallaxes were available from $HIPPARCOS$ \citep{perrymanetal1997} or $GAIA$ \citep{gaiacollaboration2016a, gaiacollaboration2016b}, they were also included.
\end{enumerate}

With all this input, \vespa\ makes a representative population for each false-positive scenario (and the transiting planet scenario). False-positive scenarios considered by \vespa\ include a background (or foreground) eclipsing binary blended with the target star, a hierarchical eclipsing binary system, or a single eclipsing binary system (as well as each of those scenarios with twice the period). Then \vespa\ calculates a prior for each scenario by multiplying together the probability that the scenario exists within the photometric aperture, the geometric probability that the scenario leads to an eclipse, and the fraction of instances in which the eclipse is appropriate. Here ``appropriate'' means that an instance's eclipse takes place within the photometric aperture, the secondary eclipse (if it exists) is not deep enough to cross the detection threshold, the instance's primary star has wavelength-dependent magnitudes within 0.1 mag of those provided as input, and the instance's primary star has $T_{\mathrm{eff}}$ and $\log{g}$ that agree with the input values to within $3\sigma$.

Then the likelihood for each scenario is calculated using the folded light curve by fitting against a distribution of transit durations, depths, and ingress/egress slopes calculated from each scenario. When the priors and likelihoods have been determined for each scenario, the last step is simply to combine them for each scenario in order to determine an overall posterior likelihood for every scenario. If the posterior likelihood for the planet scenario is exceedingly higher than all of the other scenarios combined, then the candidate can be classified as a validated planet. In our case, we required the sum of all false-positive scenario probabilities to be $<0.001$.

The output from \vespa\ includes simulation information from the underlying light-curve fitting process, as well as figures and text files conveying likelihood information of various false-positive scenarios and the transiting planet scenario. Section \ref{EPIC212521166} provides a concrete validation example with characteristic input and output. Additionally, the full \vespa\ input and output for each exoplanet candidate has been archived at \url{\texttt{https://zenodo.org/record/1164791}}.

As we noted in Section \ref{stellar_parameters}, for systems where we had collected more than one good TRES spectrum, we phased the RVs to search for or eliminate the possibility of an eclipsing binary scenario. (Small RVs would not rule out hierarchical or background eclipsing binary scenarios since large RVs would be reduced by the third star in the aperture.) In cases where the eclipsing binary scenario could be eliminated, we reduced the probability of the scenario (and the similar scenario analyzed by \vespa\ with double the period) to zero. We then divided the probability of all the remaining scenarios investigated by \vespa\ by their sum so that the total probability was still one.

Systems with multiple planet candidates are more likely to be hosting multiple planets than multiple false-positive signals. In fact, the likelihood of the planet scenario for each individual candidate is consequently boosted relative to false-positive scenarios in multiplanet candidate systems \citep{lathametal2011}. To account for this effect, we apply a ``multiplicity boost'' to the planet scenario prior in such systems, deflating the FPP by the multiplicity boost factor to account for the nature of these systems. We choose a boost factor of 25 for double-candidate systems and a boost factor of 50 for systems with three or more candidates based on the values used by \citet{lissaueretal2012}, \citet{vanderburgetal2016b}, and \citet{sinukoffetal2016}. The FPP values reported in Table~\ref{table:candidate_parameters_table} already have the appropriate multiplicity boost factor applied.

\subsection{Criteria for Planet Validation} \label{vespa_reporting}

Following standard practice, we only validated planets for which the following were true, since \vespa\ assumes we have checked for and ruled out each of them:
\begin{enumerate}
    \item There is not a composite spectrum.
    \item There is not a companion in the aperture. (This was determined from archival imaging.)
    \item There is not a companion in AO and/or speckle images. (This was determined from our AO and speckle images as well as from any images for our candidates uploaded to https://exofop.ipac.caltech.edu/k2/ before 2017 August 9.)
    \item There is no evidence of a secondary eclipse (at any arbitrary phase).
    \item There is not an ephemeris match (see Section \ref{candidates_and_vetting}).
\end{enumerate}

If any of the above were false, the FPP was not reported in Table~\ref{table:candidate_parameters_table} and the candidate was not validated regardless of the FPP value \vespa\ provided.

It is worth noting that six validated planets have recently been shown to be false positives by follow-up observations and analysis \citep{cabreraetal2017, shporeretal2017}. We briefly examine whether these instances are a cause for concern in our own results.

In the case of \citet{cabreraetal2017}, they found that two false positives exhibited increased transit depths for larger aperture masks (suggesting that a nearby star was responsible for the eclipses), while the third showed a secondary eclipse at phase = $0.62$ when they used a different data reduction process. The first two false positives would have failed the variable-sized mask test we apply via our diagnostic plots, while the secondary eclipse from the third false positive would have been caught through a combination of our data reduction process and the model-shift uniqueness test we apply \citep{coughlinetal2016}.

As for \citet{shporeretal2017}, the \vespa\ fit to the available photometry for each false positive was very poor and all three of the false positives were reported to be very large ``planets.'' We addressed these concerns in two ways. First, we decided to only use 2MASS photometry in order to avoid systematic errors between photometric bands. Second, we chose not to validate any planet candidates larger than 8 $R_\oplus$, since a candidate of that size or larger could actually be a small M dwarf. The only exception to this rule is if there are RV measurements that can rule out the eclipsing binary scenario, in which case we do report the FPP (and validate for FPP $< 0.001$), regardless of planet size.

\begin{table*}[ht!]
\begin{center}
\caption{System and planetary parameters of EPIC 212521166 \label{table:EPIC212521166_table}}
\begin{tabular}{llcc}
\tableline
\tableline
Parameter & Unit & This Paper & \citet{osbornetal2017}\\
\tableline
\multicolumn{4}{l}{\textit{Orbital parameters}} \\\\

Period $P$ & days & $13.86391^{+0.00022}_{-0.00023}$ & $13.86375 \pm 2.6{\times}10^{-4}$\\
Time of first transit\tablenotemark{a} & BJD-2454833 & $2386.87440^{+0.00072}_{-0.00068}$ & $2442.32992 \pm 6.1{\times}10^{−4}$ \\
Orbital eccentricity $e$ & ... & 0 (fixed) & $0.079 \pm 0.07$\\
Inclination & degrees & $89.36^{+0.43}_{-0.75}$ & $89.35^{+0.41}_{-0.24}$\\\\

\tableline
\multicolumn{4}{l}{\textit{Transit parameters}} \\\\

System scale $a/R_*$ & ... & $32.3^{+2.0}_{-5.7}$ & $30.8 \pm 1.0$ \\
Impact parameter $b$ & ... & $0.36^{+0.28}_{-0.23}$ & $0.34^{+0.14}_{-0.22}$ \\
Transit duration $T_{14}$ & hr & $3.181^{+0.072}_{-0.047}$ & $3.22 \pm 0.03$ \\
Radius ratio $R_\mathrm{p}/R_*$ & ... & $0.0334^{+0.0018}_{-0.0007}$ & $0.0333 \pm 6.6{\times}10^{-4}$ \\\\

\tableline
\multicolumn{4}{l}{\textit{Planet parameters}} \\\\

Planet radius $R_\mathrm{p}$ & $R_\oplus$ & $2.52^{+0.16}_{-0.10}$ & $2.592 \pm 0.098$ \\

\tableline
\multicolumn{4}{l}{\textit{Stellar parameters}} \\\\

Stellar mass $M_*$ & $M_\odot$ & $0.724 \pm 0.025$ & $0.738 \pm 0.018$ \\
Stellar radius $R_*$ & $R_\odot$ & $0.692^{+0.025}_{-0.023}$ & $0.713 \pm 0.020$ \\
Effective temperature T$_{\mathrm{eff}}$ & K & $4877 \pm 50$ & $5010 \pm 50$ \\
Surface gravity $\log{g}$ & g cm$^{-2}$ & $4.51 \pm 0.10$ & $4.60 \pm 0.03$ \\
Metallicity {[m/H]\tablenotemark{b}} & dex & $-0.30 \pm 0.08$ & $-0.34 \pm 0.03$ \\\\

\tableline
\multicolumn{4}{l}{\textit{Validation parameters}} \\\\

FPP & ... & $8.44{\times}10^{-7}$ & - \\

\tableline
\end{tabular}
\tablenotetext{a}{Our reported transit time and that reported by \citet{osbornetal2017} differ by four orbital periods.}
\tablenotetext{b}{Our reported metallicity is [m/H] (derived from many metal absorption features), while the metallicity reported in \citet{osbornetal2017} is [Fe/H] (derived from iron absorption lines only).}
\end{center}
\end{table*}

\section{Results} \label{results}

The results section is divided into two parts. In the first part, the process of validation is described in detail for a single planet candidate, in order to explain precisely how the photometry and follow-up observations are used. In the second part, our validation process is more widely applied to every selected candidate, and the results for each candidate are reported.

\subsection{Validating a Single Planet: EPIC 212521166.01} \label{EPIC212521166}

\begin{figure*}[h!]
\epsscale{1.0}
  \begin{center}
      \leavevmode
\plotone{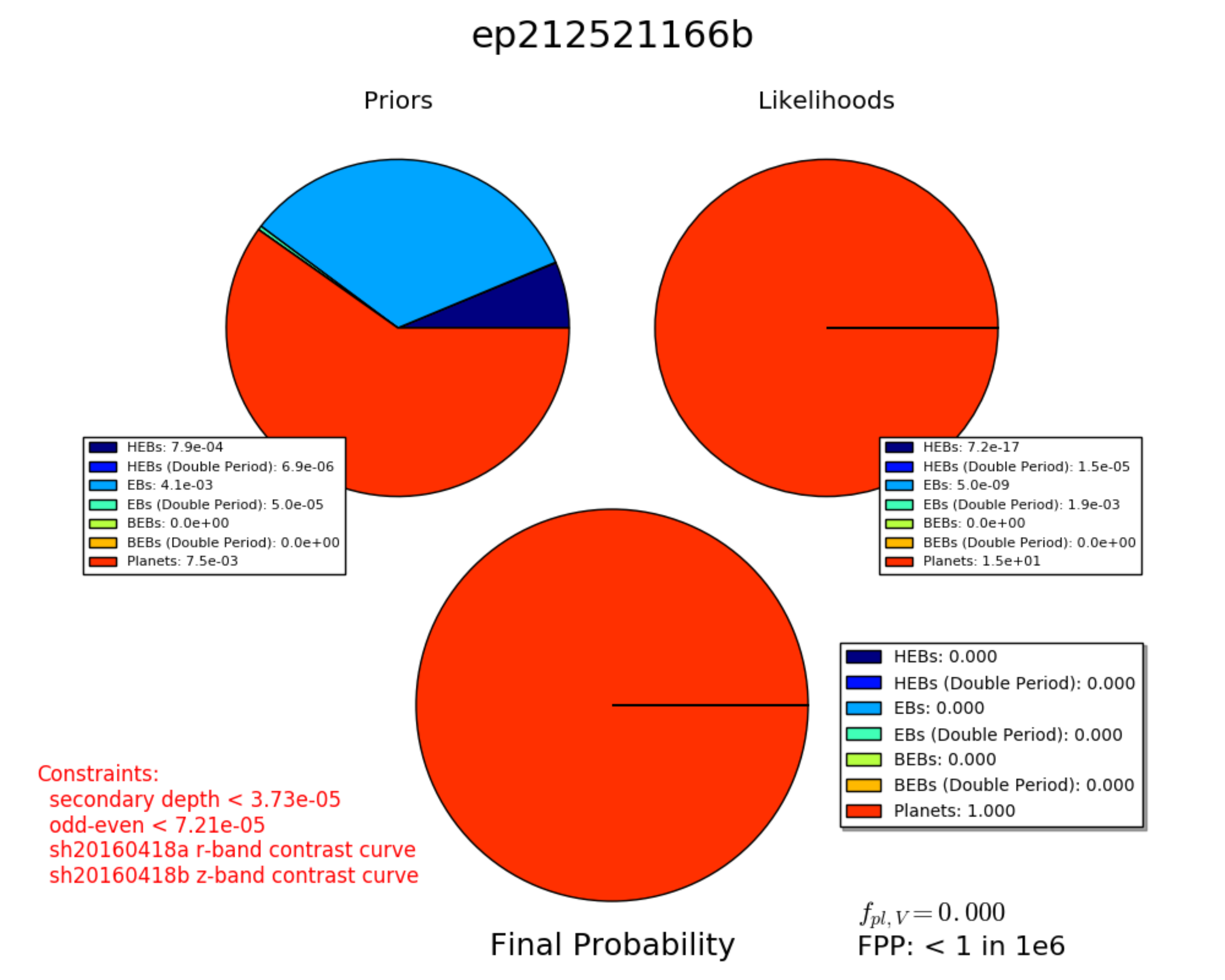}
\caption{False-positive Probability analysis of EPIC 212521166.01. HEB, EB, and BEB refer to a hierarchical eclipsing binary, eclipsing binary, and background eclipsing binary scenario, respectively. Combining the prior likelihood of a false-positive scenario (given sky position, contrast curve data, and wavelength-dependent magnitudes), as well as the likelihood of the transit photometry under various scenarios, the posterior distribution highly favors the planet scenario, with FPP = $8.44{\times}10^{-7}$. (Note: the true FPP value is always reported in a supplementary file, but for FPP $<1$ in $1$e$6$, the figure produced by \vespa\ simply reports the FPP as $<1$ in $1$e$6$.)} \label{EPIC212521166_vespa}
\end{center}
\end{figure*}

In order to understand the process that was applied to each of our candidates, it is instructive to look at validation for a single, concrete example. Here, we describe in detail the validation process for a typical planet candidate system, EPIC 212521166 (K2-110). We chose this system because (1) it was detected in our pipeline, and (2) \citet{osbornetal2017} have already confirmed and characterized the planet, which allows us to compare some of the system parameters we calculated with their results.

EPIC 212521166 was observed by \Ktwo\ during Campaign 6. After we produced a light curve from calibrated \Ktwo\ pixel data and removed systematics (as described in Section \ref{k2_data_reduction}), we searched for transits in our processed light curve (as described in Section \ref{tces_and_triage}). Our transit search pipeline detected a single TCE with a high S/N, a period of 13.87 days and a depth of about 0.1\%. The signal passed triage (see Section \ref{tces_and_triage}) and moved on to vetting (see Section \ref{candidates_and_vetting}). We confirmed that the signal was transit-like, and observed no significant secondary eclipse, phase variations, differences in depth between odd and even transits, aperture-size dependent transit depths, or other indications of a false positive or instrumental false alarm (see Figs.~\ref{vetting1}, \ref{vetting2}, and \ref{modelshift}). As a result, we promoted the EPIC 212521166 TCE to a planet candidate.

Upon its identification as a planet candidate, we observed EPIC 212521166 with the TRES spectrograph on the Mt. Hopkins 1.5m Tillinghast reflector and with the DSSI speckle-imaging camera on the 3.5 m WIYN telescope at Kitt Peak National Observatory. Using the orbital and transit parameters determined with our transit model, the stellar parameters derived from SPC, a folded light curve of the planetary transit, and two contrast curves in $r$ band and $z$ band collected via high-contrast speckle imaging from DSSI on the WIYN telescope (see Fig.~\ref{EPIC212521166_cc}), \vespa\ was employed to determine the FPP for EPIC 212521166.01. The FPP was found to be $8.44{\times}10^{-7}$, which was well below the cutoff threshold, so the planet candidate was classified as validated. The key output figure of \vespa\ can be seen in Fig.~\ref{EPIC212521166_vespa}.

Like us, \citet{osbornetal2017} found EPIC 212521166 to be a metal-poor K-dwarf star hosting planet candidate with $P$ = $13.9$ day and $R_\mathrm{p}$ = $2.6$ $R_\oplus$. A comparison of planetary and system parameters can be see in Table \ref{table:EPIC212521166_table}. Our analyses and theirs are in good agreement for all parameters. Additionally, \citet{osbornetal2017} took the further step of obtaining precise RV observations to {\em confirm} the existence of EPIC 212521166.01, so we can be confident that in this case, the assessment by \vespa of a low FPP was well justified. 

\subsection{Full Validation Results} \label{full_results}

The process of validation described for EPIC 212521166.01 in the previous section was similarly applied to the remaining candidates suitable for validation. We identified \ntrescands\ candidates in \ntrescandsys\ systems that had at least one usable TRES spectrum, and the FPP was calculated for each of these candidates (see Table~\ref{table:candidate_parameters_table}).

Occasionally, \vespa\ failed to return an FPP; in such cases, the lowest data point in the light curve was removed (to aid the initialization process for the \vespa trapezoidal transit fit), and \vespa\ was rerun. This approach worked in most cases, but if it failed, the lowest two data points were removed and \vespa\ was rerun. If that was also unsuccessful, then the FPP was not reported. (Most of the time, \vespa\ only failed after these steps because of a Roche lobe overflow error.) \nplanets\ candidates in \nsys\ systems had an FPP $< 0.001$ and were thus promoted to validated planet status.

To date, the largest single release of \Ktwo\ validation results has been \citet{crossfieldetal2016}, with 197 candidates and 104 validated planets in C0-C4. In comparison, 108 of our candidates are from C0-C4, 69 of which are validated. The two samples share 53 candidates in common, 37 of which are validated and 9 of which remain candidates in both analyses. \citep[Additionally, 2 candidates in common are only validated in this work, while 5 are only validated in ][]{crossfieldetal2016}. This leaves 55 candidates in our C0-C4 sample (30 of which are validated) that were undetected by \citet{crossfieldetal2016}, as well as 146 candidates (62 of which are validated) in the \citet{crossfieldetal2016} sample that are undetected in our own C0-C4 sample. Only ${\sim}21\%$ of the total candidates were detected by both analyses, and only ${\sim}26\%$ of the total validated planets were validated by both analyses.

% Based on table from https://arxiv.org/pdf/1703.07416.pdf
\tabletypesize{\normalsize} 
\LongTables
\begin{deluxetable}{c|cc|c}
\tabletypesize{\footnotesize} 
\tablecolumns{4} 
\tablewidth{0pt} 
\tablecaption{Breakdown of Candidate Dispositions\label{table:disp}}
\startdata
\hhline{====}
\multicolumn{1}{c}{\rule{0pt}{4ex}Previous} & \multicolumn{3}{c}{Updated Disposition}\\
\cline{2-4}
\multicolumn{1}{c}{\rule{0pt}{2ex}Disposition\tablenotemark{1}} & \multicolumn{1}{c}{VP} & \multicolumn{1}{c}{PC} &
\multicolumn{1}{c}{All} \\
\hline
 \rule{0pt}{2ex}
 VP & \nalreadyplanets & 15 & 68 \\
 PC & \nalreadycand & 55 & 94 \\
 FP & 1\tablenotemark{2} & 1\tablenotemark{3} & 2 \\
 UK & \nnew & 55 & 111 \\
 \hline
 \rule{0pt}{2ex}All & 149 & 126 & 275
\enddata
\tablenotetext{1}{VP = Validated Planet, PC = Planet Candidate, FP = False Positive, UK = Unknown}
\tablenotetext{2}{EPIC 210894022.01. See Section \ref{full_results}}
\tablenotetext{3}{EPIC 202900527.01. See Section \ref{full_results}}
\end{deluxetable}

The sample overlap may seem surprisingly small, but it makes more sense when the candidate selection and validation processes are examined. For example, \citet{crossfieldetal2016} only considered candidates with $1$ day $<$ P $< 37$ day ($19\%$ of our C0-C4 sample was outside that range), and we only considered candidates with $K_\mathrm{p} < 13$ ($48\%$ of their sample was outside that range). If we only consider C0-C4 candidates within those ranges (137 total), both teams find over half of each other's samples, and the overlap between samples rises to $39\%$ (53 candidates). Similarly, for validated planets within these ranges (77 total), both teams find more than two-thirds of each other's samples, and the overlap between samples rises to $57\%$ (44 validated planets).

There are many further examples of differences that created discrepancies between the two samples. We required that each planet candidate had at least one usable TRES spectrum (see Sections ~\ref{TRES} and ~\ref{stellar_parameters}), even though in early campaigns, TRES spectra were not collected for all bright candidates. This led to the exclusion of many otherwise promising candidates that are included in \citet{crossfieldetal2016}. Further discrepancies could have arisen from the temperature and planet radius cuts applied to our validated planet sample, the elimination of objects with companions closer than 4'' in the \citet{crossfieldetal2016} candidate sample, and a difference in significance thresholds for TCEs (we used $9\sigma$ while they used $12\sigma$).

Table~\ref{table:disp} compares the candidate dispositions found in this work with their previous dispositions (according to the NASA Exoplanet Archive\footnote{\url{https://exoplanetarchive.ipac.caltech.edu/}} and the Mikulski Archive for Space Telescopes\footnote{\url{https://archive.stsci.edu/k2/published\_planets/search.php}}; both accessed 2018 February 14). We should note that 15 of our candidates were not validated in this work even though they have been previously validated elsewhere. However, all of these candidates were either restricted from being validated by our conservative criteria for validation (see Section~\ref{vespa_reporting}), or had an FPP value close to our validation cutoff of FPP = 0.001, or were validated using different or additional observations as input to \vespa. We also note that our work classifies two targets that have previously been labeled as false positives: EPIC 202900527.01 (K2-51 b) and EPIC 210894022.01 (K2-111 b). \citet{shporeretal2017} clearly showed EPIC 202900527.01 to be a stellar binary. We do not claim otherwise by labeling it a candidate (since we label all targets with  FPP $> 0.001$ as candidates). On the other hand, \citet{crossfieldetal2016} previously identified EPIC 210894022.01 as a false positive, but a subsequent, improved \vespa\ run showed the target to in fact be a planet (I. Crossfield 2017, private communication). Therefore we do claim this target to be validated.

Another case worth mentioning is the multiplanet system EPIC 228725972, which hosts one validated planet and one candidate ruled out by a companion in the aperture and in a NESSI speckle image. The smallest aperture mask we tested excluded the companion (located approximately 12 arcsec from the primary), but only exhibited transits for one of the candidates, hence the difference in flags for the two candidates.

In addition to the FPPs and other parameters derived through the validation process described in previous sections, we also calculated the radii and masses of the stars in our sample (the stellar radius allowed us to calculate the absolute planetary radius). We calculated these parameters by inputting stellar effective temperature, metallicity, and surface gravity into the \texttt{isochrones} Python package \citep{morton2015a}. We also found that inputting 2MASS photometry into \texttt{isochrones} along with the aforementioned stellar parameters had no noticeable effect on the resulting stellar radius and mass values (or their uncertainties). Although isochrones have been found to underestimate stellar radii before \citep{dressingetal2017a, martinezetal2017, dressingetal2017b}, this effect is confined to M dwarfs and late-K dwarfs, which are largely excluded from our host star sample since we do not consider stars with effective temperatures below 4250 K (see Section~\ref{stellar_parameters}). The derived stellar masses and radii are reported in Table~\ref{table:candidate_parameters_table}. (We also note that all stellar parameters for each system are reported in Table~\ref{table:stellar_parameters_table}.)

\section{Discussion} \label{discussion}

\subsection{Lessons Learned} \label{lessons_learned}

There have been a few recent instances of false-positive misclassification \citep{cabreraetal2017,shporeretal2017}, in which a target has been classified as a validated planet but was later shown to be a false positive through subsequent analysis. Here, we discuss some of the pitfalls of statistical validation, and share some lessons we have learned and solutions we have implemented to prevent false-positive misclassification.

First, spectroscopy is key. Stellar parameters are crucial to understanding the host star and correctly classifying the candidates. For example, identifying the star as highly evolved can prevent an orbiting brown dwarf or M dwarf from being incorrectly labeled as a smaller planet orbiting a main-sequence star. One might attempt to avoid this issue by estimating stellar surface gravity via photometry alone, but that is notoriously difficult to do and can lead to incorrectly estimated stellar parameters. Collecting at least one spectrum from which to estimate spectroscopic parameters (including surface gravity) can avoid many issues like this, and collecting more spectra may exclude (or reveal) large RV variations from an eclipsing binary scenario.

Second, a search for secondary eclipses at \textit{all} phases should be performed to exclude a scenario with a highly eccentric binary. The search method used here was the model-shift uniqueness test designed by \citet{coughlinetal2016}. This method compared off-transit portions of the light curve with the transit portion of the best-fit transit model (and the transit model inverted, to provide a visual comparison of detection significance). An even more rigorous method would consider secondary eclipses of duration and depth distinct from that of the transit.

Third, when validating exoplanets using \vespa\, one should be very careful about which broadband photometry is used and what offsets and systematics there are. Statistical validation work on data from the original \Kepler\ mission took advantage of high-quality and uniform broadband photometry from the \Kepler\ Input Catalog \citep[KIC,][]{brownetal2011}. The analogous EPIC catalog for \Ktwo\ is similarly vital to the analysis of the \Ktwo\ star and planet sample. However, it is approximately four times larger and was produced by collecting archival photometric measurements; therefore, it is more heterogeneous and more prone to systematic errors and offsets than the KIC. We have found that in some cases, these systematic errors or offsets in photometry can result in untrustworthy FPP estimates, in particular, extremely low FPP measurements for known or likely false positives. For example, when we input to \vespa\ all available broadband photometry from the EPIC catalog for the candidate EPIC 211418290.01, \vespa\ yielded an FPP lower than $10^{-6}$ because it was unable to find an acceptable fit to a binary star model due to underestimated photometric uncertainties and therefore incorrectly ruled this false-positive scenario out. However, inputting only 2MASS photometry to \vespa\ yielded a better fit to the eclipsing binary scenario, and \vespa\ therefore returned a much higher FPP of about $0.6$. Situations like this were not uncommon, which is why we ultimately decided to use only photometry from the all sky 2MASS survey. We chose to use only the 2MASS survey because it provides broadband photometry for all of our candidates and operates in the infrared, reducing the effects of reddening. Including photometry from additional surveys that are calibrated differently and have distinct systematic uncertainties would introduce non-uniformities that would be difficult to quantify and could bias our results.

To be conservative, we also added 0.02 mag of systematic error in quadrature with the reported error for each 2MASS band we used. This will continue to be important after the launch of $TESS$, which will observe targets selected from the $TESS$ Input Catalog \citep[TIC,][]{stassunetal2017}. The TIC will be very important for selecting candidates for observation and follow-up; however, it is being assembled in a manner similar to the EPIC and is approximately 10 times larger, so it may have similar or even greater photometric offsets. 

Fourth, it is very important to make and remake a candidate's light curve with multiple apertures of varying sizes, in order to determine whether the transit signal originates from the target star or a nearby background star (and thus exclude background false-positive scenarios). This test caught many false positives in our own analysis, and has been used successfully before in other candidacy analyses \citep[e.g.][]{dressingetal2017b}.

Fifth, transiting giant planets are nearly indistinguishable from eclipsing brown dwarfs or small M-dwarf stars, since they have roughly equal radii. Great care should be taken when interpreting the output of \vespa\ for giant planets, and unless there is a clear way to distinguish between the two scenarios (via RV measurements, secondary eclipse detection, composite spectra, etc.), attempts to validate large planets can be difficult and prone to error.

By considering each of the above lessons and incorporating them into future validation analyses, misclassifications of validated planets can be significantly reduced and hopefully eliminated in all but the most perverse scenarios.

We will now investigate some of the individual validated planets in our sample as well as the characteristics of our sample as a whole.

\subsection{A New Brightest Host Star in the \Ktwo\ and Kepler Planet Samples}

One interesting target in our sample is EPIC 205904628 (HIP 110758, HD 212657), an F7 star observed in Campaign 3. We detected and validated a 2.8 $R_\oplus$ planet on a 10-day orbit. At a $V$ magnitude of 8.24, EPIC 205904628 is now the brightest star at optical wavelengths in the entire \Kepler\ and \Ktwo\ samples to host a validated planet. EPIC 205904628 is just slightly brighter than \Kepler-21 ($V = 8.25$), a star of similar spectral type found during the original \Kepler\ mission to host a single short-period exoplanet \citep{howelletal2012, lopez-moralesetal2016}. $TESS$ is expected to discover planets around many stars this bright and brighter.

We note that due to its brightness, we applied some special care to this star. In particular, we re-reduced the \Ktwo\ light curve by extracting light curves from larger photometric apertures than our standard pipeline uses. We also incorporated information from 2MASS $J$-band imaging to rule out additional stars in the photometric aperture by calculating a contrast curve and inputting it to \vespa\ to provide deeper imaging constraints than our speckle image.

\subsection{Characteristics of the Sample} \label{sample_characteristics}

In addition to individual planets in our sample, it is interesting to explore a few of the demographics of the newly validated exoplanet population. Figs.~\ref{kp_hist_with_kepler}-\ref{flux_v_rp_contour} reveal various aspects of the validated exoplanet sample and the exoplanet candidate sample.

\subsubsection{The Stellar Magnitude Distribution}

\begin{figure}[h!]
\epsscale{1.15}
  \begin{center}
      \leavevmode
\plotone{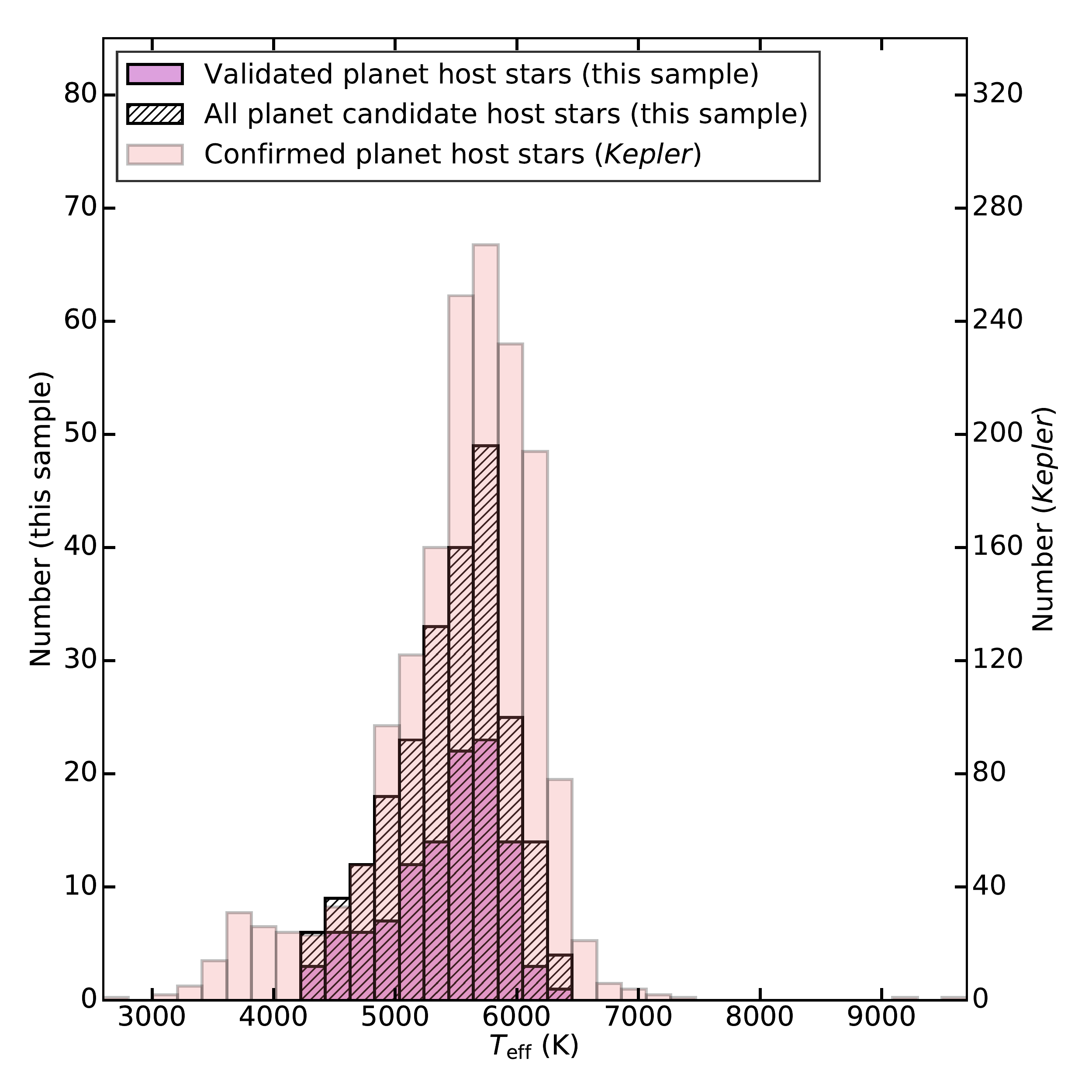}
\caption{Histograms of the Kepler magnitude for host stars to validated planets (purple) and candidate planets (diagonal lines) from C0-C10 of \Ktwo\ that have been identified in this work, as well as a comparison with host stars to confirmed Kepler planets (pink). There is a fairly sharp cutoff in our sample near magnitude 13, since validation was only conducted on stars for which we had a spectrum and almost all of the spectra we used were for targets brighter than 13th mag. The typical Kepler host star is much fainter, with the peak of the distribution near 15th mag, which makes \Ktwo\ target stars much more suitable for follow-up observations than Kepler stars. Moreover, the brightest star hosting a validated planet in our sample is EPIC 205904628, a $V$ = 8.2 star with a 2.8 $R_\oplus$ planet on a 10-day orbit. Validated in this work, EPIC 205904628 is now the brightest host star for a validated planet in either the \Kepler\ or \Ktwo\ samples.} \label{kp_hist_with_kepler}
\end{center}
\end{figure}

Fig.~\ref{kp_hist_with_kepler} is a histogram of the distribution of brightnesses for the host stars in our sample compared with the brightness for stars hosting known Kepler planets (downloaded from https://exofop.ipac.caltech.edu/; accessed 2017 December 4). Most of the candidates in our population are clustered near stellar magnitudes of $K_\mathrm{p}$ = $12$ to $K_\mathrm{p}$ = $13$, and the cutoff we imposed at $K_\mathrm{p}$ = $13$ is very evident. Fig.~\ref{kp_hist_with_kepler} also shows the distribution of brightnesses for host stars to known Kepler planets. This distribution peaks near $K_\mathrm{p}$ = $15$ and drops rapidly at brighter magnitudes. It is more difficult to find planet candidates around faint targets for \Ktwo\ than for Kepler since the baseline is shorter. Therefore \Ktwo\ candidates tend to orbit very bright host stars relative to Kepler candidates, making \Ktwo\ candidates excellent targets for follow-up observations.

\subsubsection{The Effective Temperature Distribution}

\begin{figure}[h!]
\epsscale{1.15}
  \begin{center}
      \leavevmode
\plotone{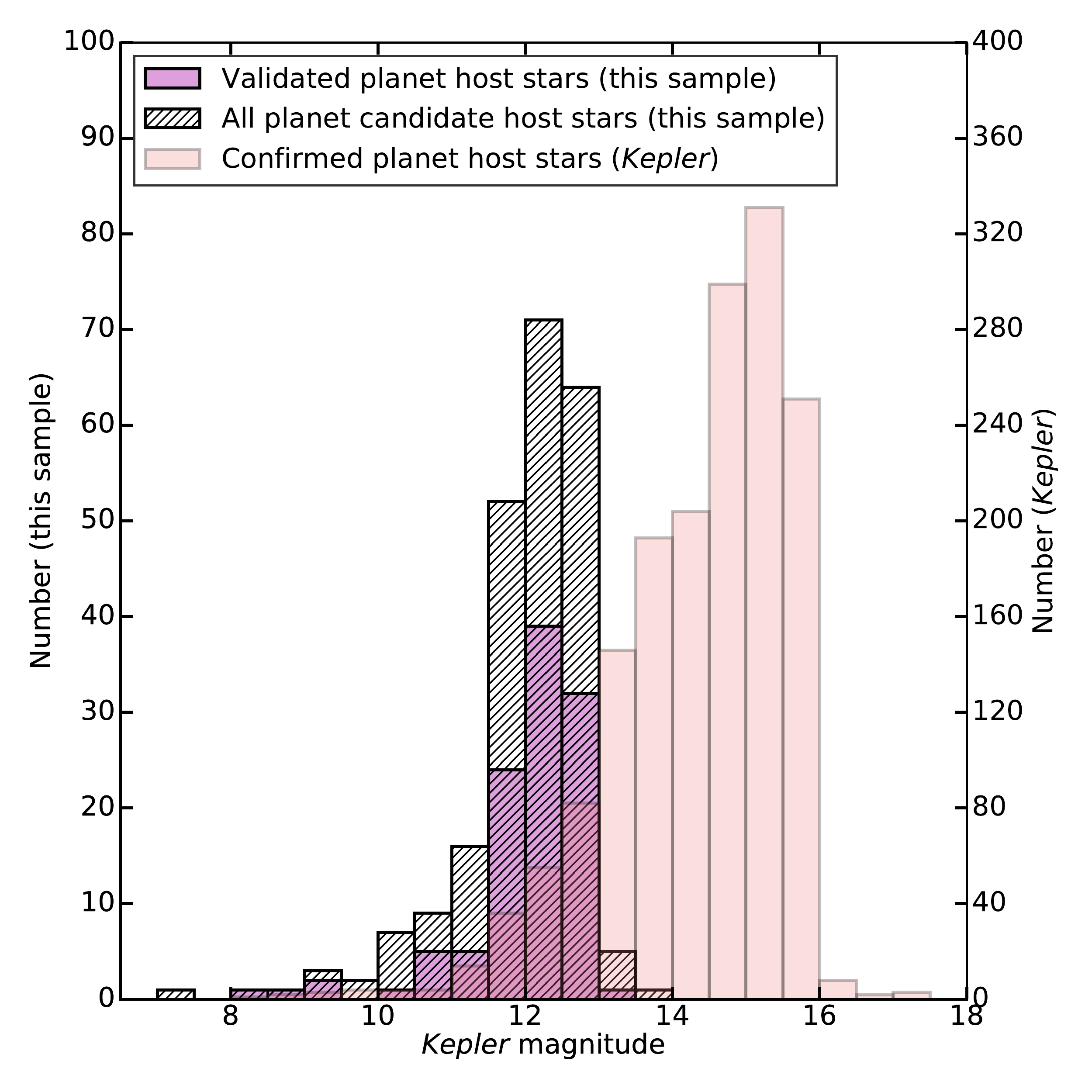}
\caption{Histograms of the effective temperature for host stars to validated planets (purple) and candidate planets (diagonal lines) from C0-C10 of \Ktwo\ that have been identified in this work, as well as a comparison against host stars to confirmed Kepler planets (pink). Except for the cutoffs we impose at $T_{\mathrm{eff}} = 4250$ K and $6500$ K, our sample follows a similar effective temperature distribution as the Kepler sample.} \label{teff_hist_with_kepler}
\end{center}
\end{figure}

Most of the validated planets and candidates we identified orbit host stars with effective temperatures in the $5000-6000$ K range. However, out of \ntrescandsys\ stars in our sample 41 are cooler than $5000$ K and 25 are hotter than $6000$ K. No stars in our sample are cooler than $4250$ K or hotter than $6500$ K. See Fig.~\ref{teff_hist_with_kepler} for a comparison with the Kepler sample (downloaded from https://exofop.ipac.caltech.edu/; accessed 2017 December 4).

\subsubsection{The Period Distribution}

\begin{figure}[h!]
\epsscale{1.15}
  \begin{center}
      \leavevmode
\plotone{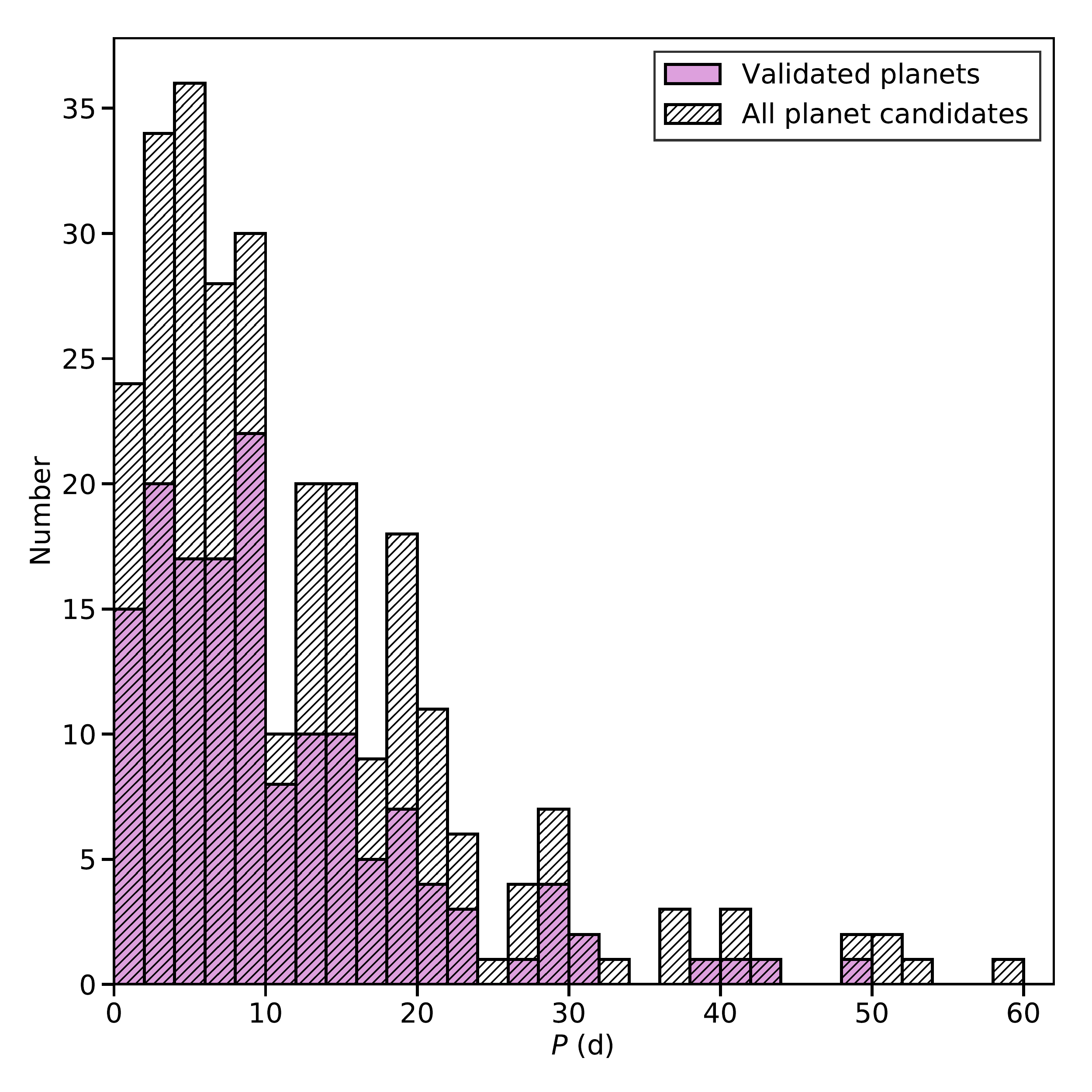}
\caption{Histogram of the orbital period for the validated planets (purple) and candidates (diagonal lines) in C0-C10 of \Ktwo\ that have been identified in this work. The steep drop-off in validated planets and candidates at around 20 days is due to the \Ktwo\ strategy of observing each field for only ${\sim}75$ days.} \label{per_hist}
\end{center}
\end{figure}

We also explored the period distribution of our candidate population. As can be seen in Fig.~\ref{per_hist}, the typical validated planet or candidate has an orbital period of 20 days or less. This should come as no surprise, given the baseline of ${\sim}75$ days of \Ktwo\ observations per campaign field. Of \ntrescands\ candidates in our sample, only 46 have periods longer than 20 days, and only 10 have periods longer than 40 days.

\subsubsection{The Multiplicity Distribution}

\begin{figure}[h!]
\epsscale{1.15}
  \begin{center}
      \leavevmode
\plotone{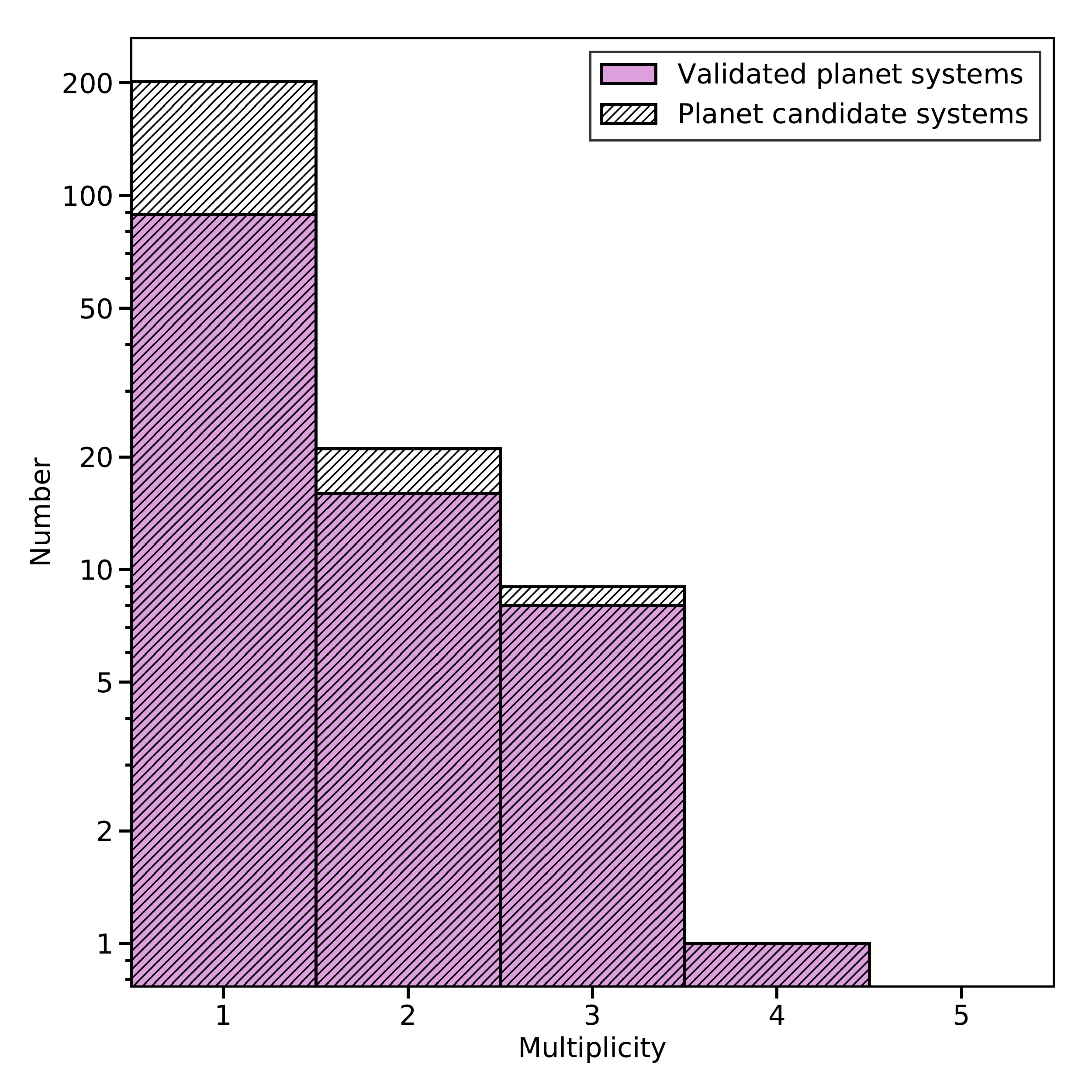}
\caption{Histogram of the multiplicity for the validated planet systems (purple) and planet candidate systems (diagonal lines) in C0-C10 of \Ktwo\ that have been identified in this work. (A validated planet system is a system for which all candidates have been validated.) Most validated planets and candidates have no detected candidate companions in their system. The largest system in the sample is EPIC 212157262, which hosts four validated planets.} \label{multiplicity_hist}
\end{center}
\end{figure}

Most of the candidates in our sample (validated or otherwise) are the only candidate in their system to have been detected. However, our sample does include 21 systems with two candidates (15 of these have only validated planets), 9 systems with three candidates (8 of those have only validated planets), and 1 system with four candidates (EPIC 212157262; all candidates in this system are validated). The full multiplicity distribution can be seen in Fig.~\ref{multiplicity_hist}.

\begin{figure*}[h!]
\epsscale{0.65}
  \begin{center}
      \leavevmode
\plotone{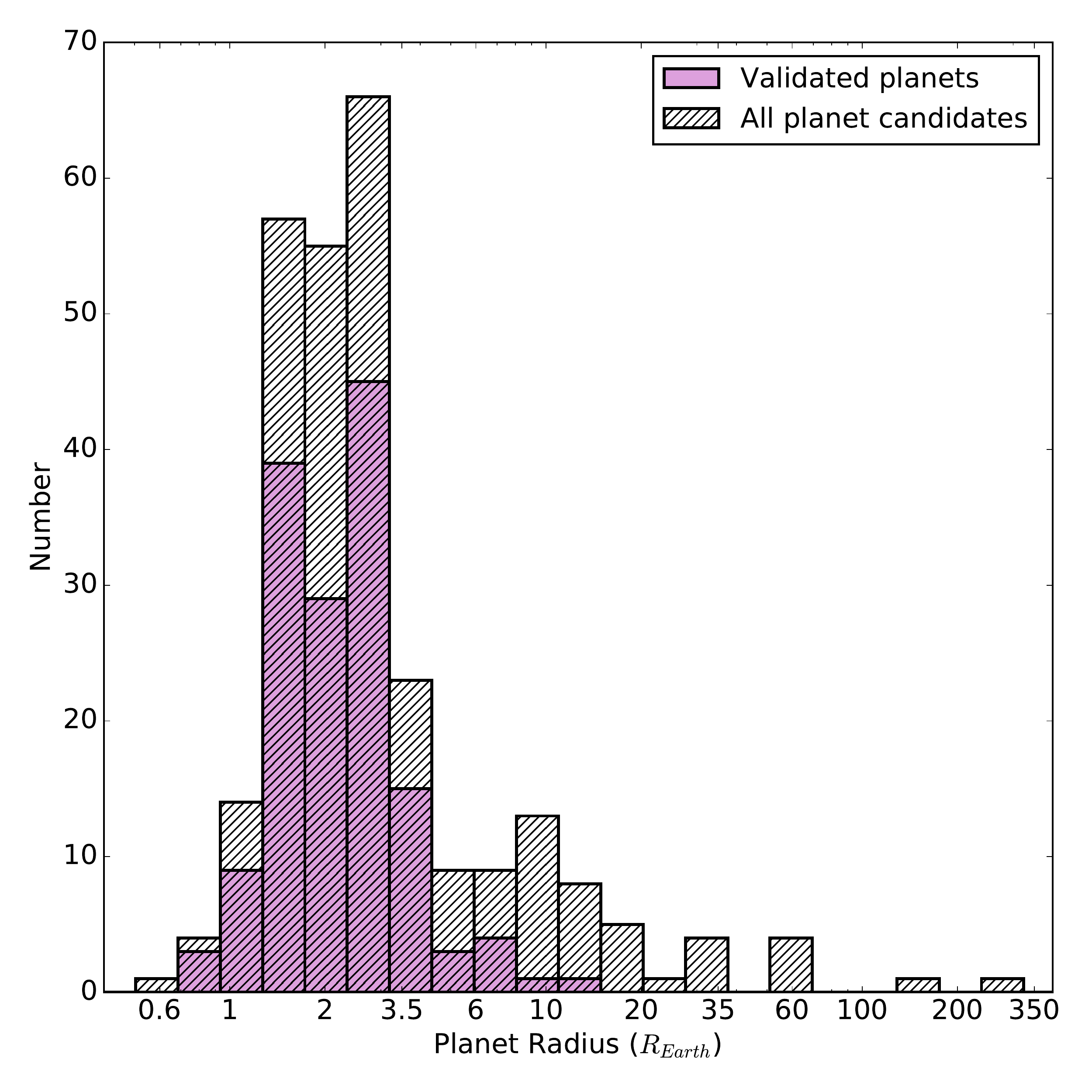}
\caption{Histogram of the planetary radius for the validated planets (purple) and candidates (diagonal lines) in C0-C10 of \Ktwo\ that have been identified in this work. See Figs.~\ref{rp_hist_uncorrected_and_corrected} and~\ref{fulton_comparison_wide_and_narrow} for narrower ranges, a completeness correction, and a comparison with the \citet{fultonetal2017} planet candidate sample.} \label{rp_hist}
\end{center}
\end{figure*}

\begin{figure*}[h!]
\epsscale{1.1}
  \begin{center}
      \leavevmode
\plotone{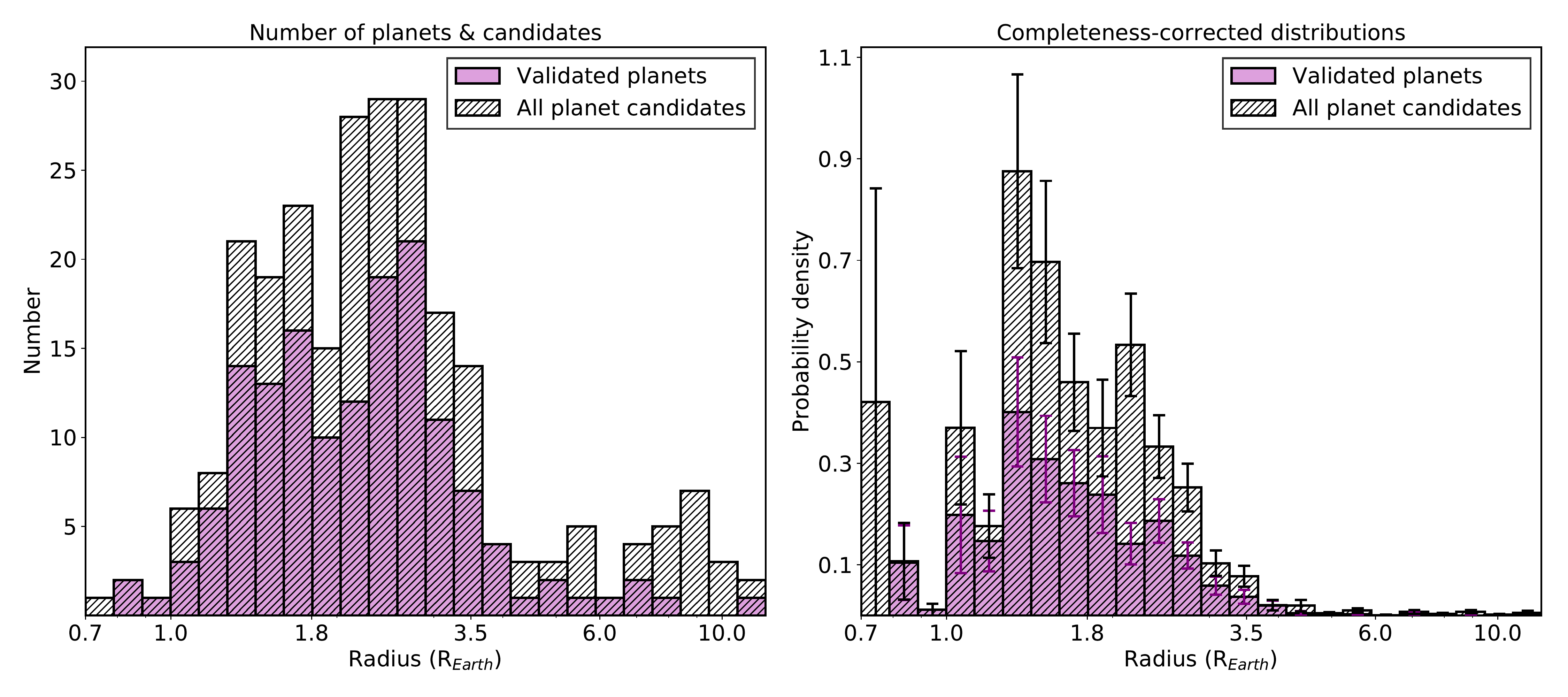}
\caption{Left: histogram of planetary radius for the validated planets (purple) and candidates (diagonal lines) in C0-C10 of \Ktwo\ that have been identified in this work (between 0.7 and 12 $R_\oplus$). Right: same data as presented in the left panel, with a completeness correction applied to estimate the underlying planet population. Error bars for each bin correspond to $1/\sqrt{n}$ (where $n$ is the number of objects in the bin) scaled according to the completeness correction subsequently performed. Through visual inspection, the full candidate sample appears to exhibit a frequency gap centered near $1.8-2.0$ $R_\oplus$.} \label{rp_hist_uncorrected_and_corrected}
\end{center}
\end{figure*}

\begin{figure*}[h!]
\epsscale{1.15}
  \begin{center}
      \leavevmode
\plotone{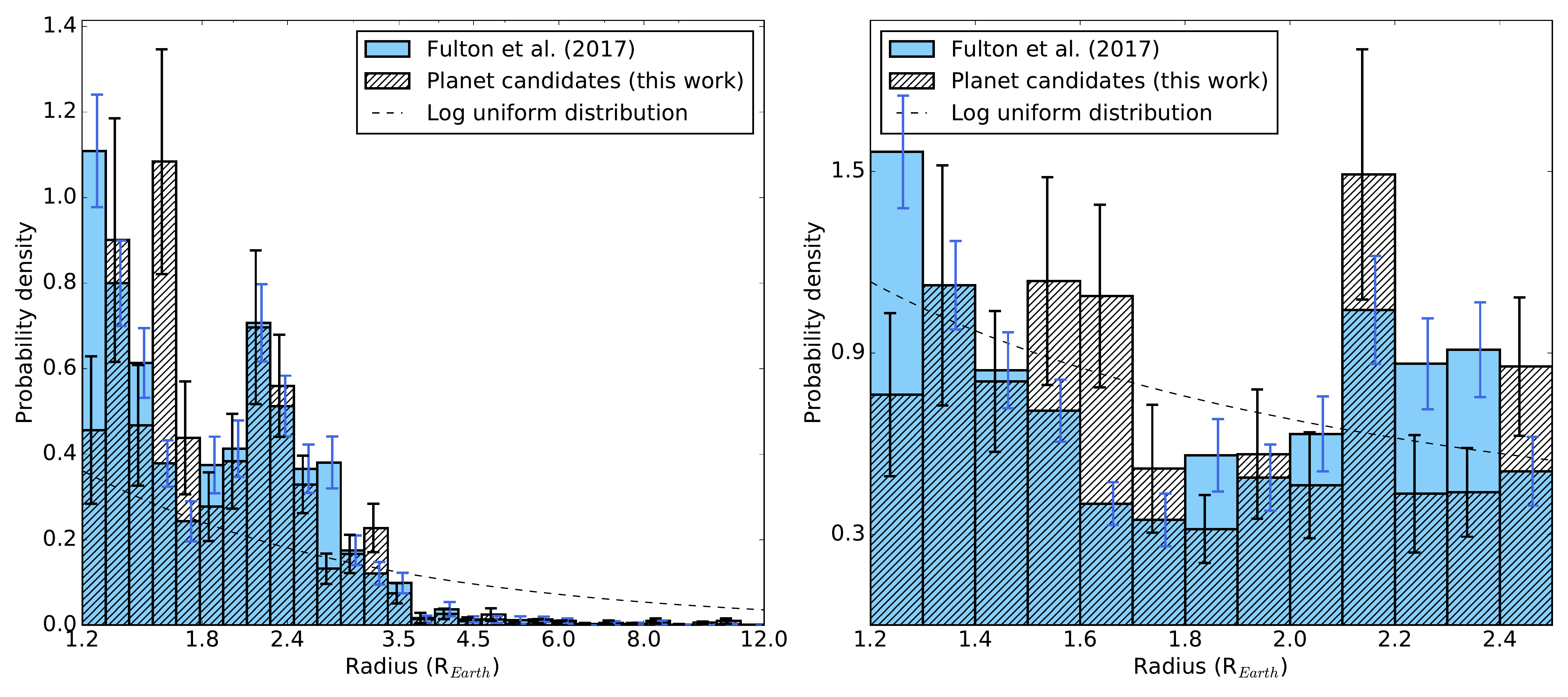}
\caption{Left: completeness-corrected comparison of the radius distribution between $1.2$ and $12.0$ $R_\oplus$ for all candidate planets identified in this work (thatched), the planet sample analyzed in \citet[][blue]{fultonetal2017}, and a normalized log-uniform distribution (dashed line). Right: comparison of the same distributions as the left panel, now focused on planet candidates with radii between $1.2$ and $2.5$ $R_\oplus$. The frequency gap identified by \citet{fultonetal2017} can clearly be seen in their distribution. A similar gap tentatively appears in our sample (near $1.8-2.0$ $R_\oplus$), although a log-uniform distribution provides a fit to our sample roughly as good as the \citet{fultonetal2017} distribution.} \label{fulton_comparison_wide_and_narrow}
\end{center}
\end{figure*}

\begin{figure*}
\centering
\parbox{9cm}{\includegraphics[width=9cm]{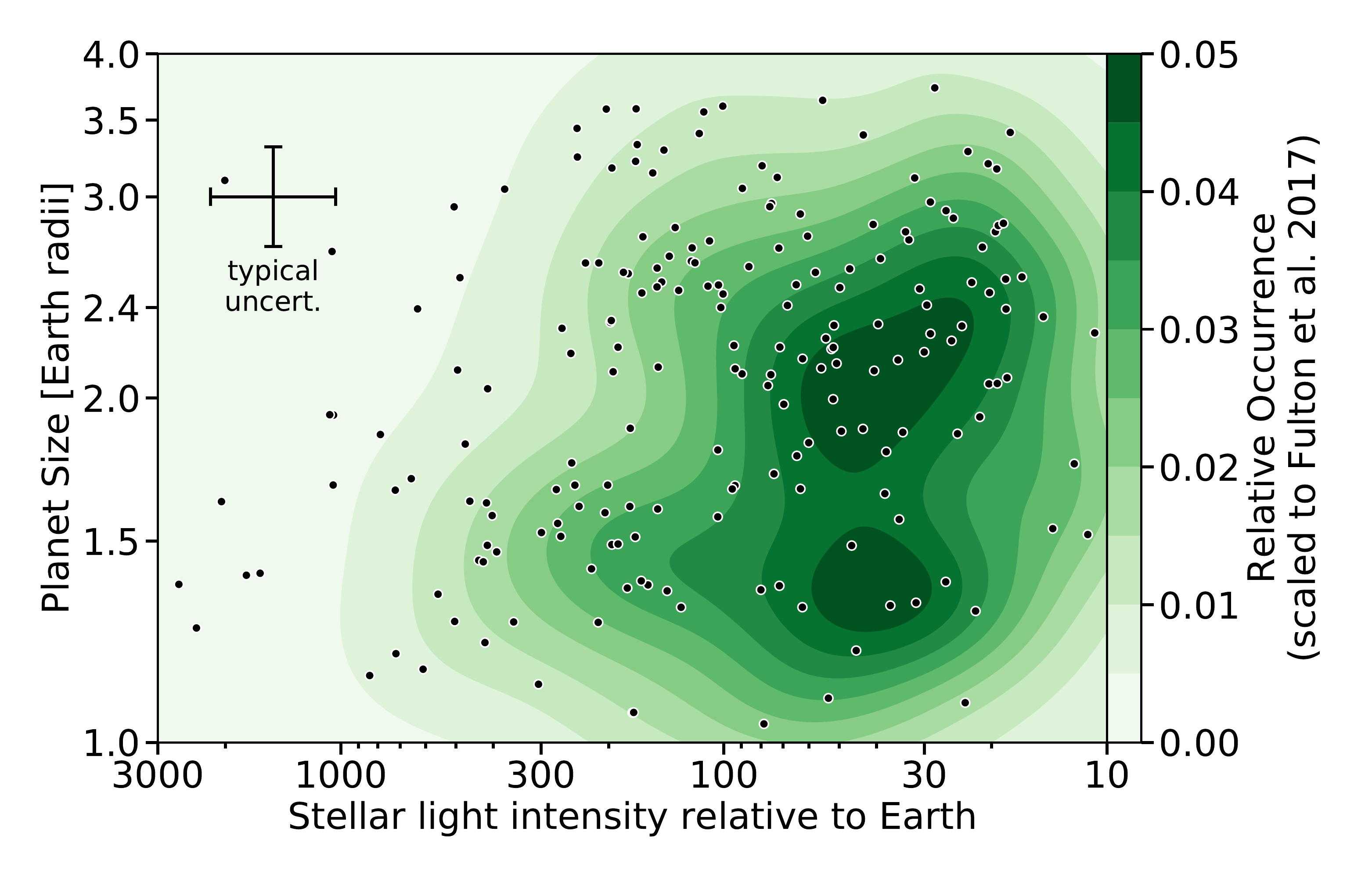}}
\qquad
\begin{minipage}{8cm}
\includegraphics[width=8cm]{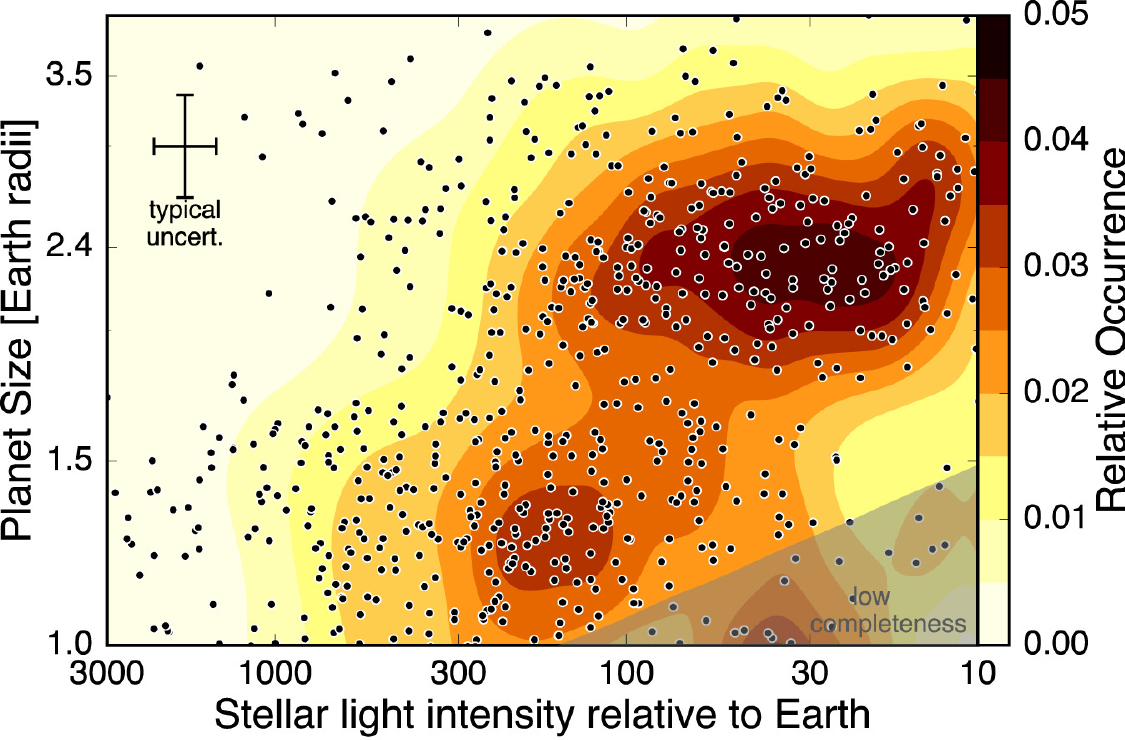}
\end{minipage}
\caption{Left: contour plot of incident stellar flux versus planetary radius. The black points are the \ntrescands\ candidates in our sample, and the typical uncertainty is plotted in the top left. The contours are completeness-corrected and roughly scaled to the same height as those in Fig.~10 from \citet{fultonetal2017}. The contours show tentative visual evidence of a gap in the radius frequency near $1.6-1.8$ $R_\oplus$. Right: top panel of Fig.~10 from \citet[][reproduced with permission]{fultonetal2017}, with a similar gap near $1.6-1.8$ $R_\oplus$.}
\label{flux_v_rp_contour}
\end{figure*}

\subsubsection{The Planet Radius Distribution}

One of the more interesting features of our sample was the radius distribution. Fig.~\ref{rp_hist} demonstrates the full planetary radius distribution of our sample for both validated planets and candidates. As expected, there is a sharp cutoff in planet radius for validated planets at $R_\mathrm{p}$ = $8$ $R_\oplus$ (since we generally chose not to validate larger planets). 

In order to investigate the radius distribution of the underlying planet population, we adopted and applied a rough completeness correction to our planet sample. We determined the completeness of our \Ktwo\ planet detection pipeline by performing injection/recovery tests. We injected planet signals into \Ktwo\ light curves over a range of expected S/Ns (analogous to the injection/recovery tests performed by \citealt{petiguraetal2013}; \citealt{dressingandcharbonneau2015}; \citealt{christiansenetal2016} on the \Kepler\ pipeline). We then calculated for each of our planet candidates the number of stars observed by \Ktwo\ around which we could have detected the candidate, and weighted the contribution of each candidate to the histograms by that factor and the geometric transit probability.

The uncorrected and corrected distributions for validated planets and candidates can be seen in Fig.~\ref{rp_hist_uncorrected_and_corrected}, while a comparison of our corrected candidate sample to the \citet{fultonetal2017} sample and a log-uniform distribution can be seen in Fig.~\ref{fulton_comparison_wide_and_narrow}. Additionally, a completeness-corrected contour plot of stellar incident flux versus planetary radius for our candidate sample can be seen in Fig.~\ref{flux_v_rp_contour} \citep[constructed in the same way as Fig.~10 from ][]{fultonetal2017}.

By visual inspection of these figures alone, our corrected candidate sample appears to exhibit a gap in the radius frequency around $1.8-2.0$ $R_\oplus$, similar to the gap in the Kepler planet sample explored by \citet{fultonetal2017}. They argued that the underlying astrophysical effect could be photoevaporation, whereby stellar incident flux strips a planet's H/He atmosphere if the atmosphere is not thick enough, leaving a population of stripped, rocky planets and an untouched population of larger, gaseous mini-Neptunes \citep{owenandwu2013, lopezandrice2016, owenandwu2017, vaneylenetal2017}. As an alternative hypothesis, they also suggested that gas accretion could be delayed during planet formation until the protoplanetary disk is already gas poor, creating a population of small, rocky planets \citep{leeetal2014, leeandchiang2016}.

We wished to analyze in a more quantitative fashion how the completeness-corrected radius distribution of our \Ktwo\ candidates compared to the completeness-corrected Kepler candidates from \citet{fultonetal2017}; we decided to also compare against a log-uniform distribution to probe the significance of a possible bimodality \citep[as did][]{fultonetal2017}. We binned all three normalized distributions in $0.1$ $R_\oplus$ intervals from $1.2$ to $2.5$ $R_\oplus$. We assigned error bars to our own sample and the \citet{fultonetal2017} sample via Poisson statistics (for the number of candidates in each bin) and then scaled the uncertainty according to the completeness-correction scaling applied to each bin.

We then calculated $\chi^2_{\rm reduced}$ between our sample and that of \citet{fultonetal2017} as well as our sample and the log-uniform distribution, finding values of $2.21$ and $2.34$, respectively. Across all bin numbers from 5 to 20 for which there were more than two candidates in every (equally sized and spaced) bin for all samples, the average $\chi^2_{\rm reduced}$ was $2.21$ between our sample and that of \citet{fultonetal2017} and $2.95$ between our sample and the log-uniform distribution. However, we note that a large portion of the $\chi^2_{\rm reduced}$ between our sample and that of \citet{fultonetal2017} is caused by only the smallest radius bin. Still, the \citet{fultonetal2017} sample did not provide a significant improvement in goodness of fit than a simple log-uniform distribution. This suggests that the visual gap in our candidate sample cannot be quantitatively confirmed unless future observations and planet detections are made to increase the sample size.

%p-values and errors for 1000 draws from 10-bin estimates of Fulton dist., our dist., and log. unif. dist.
%--------
%KS test, Fulton v. log unif: 0.0018 +/- 0.0091
%KS test, Our dist. v. log unif: 0.018 +/- 0.033
%KS test, Fulton v. Our dist.: 0.075 +/- 0.103
%--------
%(D, p-value) for 1-3 Earth radii (best radii in my data set):
%KS test, Fulton v. log unif: (0.082, 0.00045)
%KS test, Our dist. v. log unif: (0.18, 0.0017)
%KS test, Fulton v. Our dist.: (0.16, 0.013)

\subsubsection{Future Applications}

The \Ktwo\ mission has thus far conducted observations through more than 15 campaigns, and it will continue observations for future campaigns until it expends all of its fuel. Given that the validation infrastructure built for this research is directly applicable to future campaigns, we will be able to identify and validate exoplanet candidates from upcoming campaigns as quickly as the necessary follow-up observations can be collected.

This research will also be useful even after the end of the \Ktwo\ mission. The upcoming $TESS$ mission \citep{rickeretal2015} is expected to yield more than 1500 total exoplanet discoveries, but it is also estimated that $TESS$ will detect over 1000 false-positive signals \citep{sullivanetal2015}. Even so, one (out of three) of the level-one baseline science requirements for $TESS$ is to measure the masses of 50 planets with $R_\mathrm{p} < 4$ $R_\oplus$. Therefore, an extensive follow-up program to the primary photometric observations conducted by the spacecraft is required, including careful statistical validation to aid in the selection of follow-up targets. The work presented here will be extremely useful in that follow-up program, since only modest adjustments will allow for the validation of planet candidate systems identified by $TESS$ rather than \Ktwo.

\section{Summary and Conclusions} \label{conclusion}

In this paper, we processed \ntargets\ targets from C0-C10, removed instrumental systematics from the \Ktwo\ photometry, and searched for TCEs. We identified \ntces\ TCEs, which were subjected to triage and vetting. Of these, \ncands\ target systems passed both procedures and were upgraded to candidates. Of these, \ntrescands\ candidates in \ntrescandsys\ systems had at least one usable TRES spectrum. For each candidate, we also collected and analyzed follow-up observations, including spectroscopy and high-contrast imaging. We derived transit/orbital parameters from the \Ktwo\ photometry, stellar parameters from the spectra, and contrast curves from the high-contrast imaging. Then, using these results, we determined the FPP of each planet candidate using the \vespa\ validation procedure \citep{morton2012, morton2015b}, which calculates the likelihood of various false-positive scenarios as well as the true-positive scenario. These FPP values were then adjusted appropriately for systems with radial velocity measurements and for multiplanet systems (see Section \ref{vespa_application}). We reported the resulting FPPs for our planet candidates (see Table~\ref{table:candidate_parameters_table}) and classified candidates with FPP $< 0.001$ as validated planets. 

Of the \ntrescands\ candidates, \nplanets\ had an FPP below the validation threshold of $0.001$, while \nunvalidated\ remained candidates (FPP $> 0.001$). According to the NASA Exoplanet Archive\footnote{\url{https://exoplanetarchive.ipac.caltech.edu/}} and the Mikulski Archive for Space Telescopes\footnote{\url{https://archive.stsci.edu/k2/published\_planets/search.php}} (both accessed 2018 February 14), of the \nplanets\ newly validated exoplanets, \nnew\ have not been previously detected, \nalreadycand\ have already been identified as candidates, \nalreadyplanets\ have already been validated, and \nfp\ has previously been classified as a false positive (EPIC 210894022.01; see Section \ref{full_results}). As a result, this work will increase the validated \Ktwo\ planet sample by nearly 50\% (and increase the \Ktwo\ candidate sample by ${\sim}20\%$. The full disposition results can be found in Table~\ref{table:disp}.

Most of the newly validated exoplanets orbited host stars with $12 < K_\mathrm{p} < 13$ and $5000$ K $< T_{\mathrm{eff}} < 6000$ K. Additionally, the majority of validated planets had orbital periods $< 20$ days and planetary radii between 1 and 4 $R_\oplus$. Our complete candidate sample also shows signs of a frequency gap in the radius distribution (see Fig~\ref{fulton_comparison_wide_and_narrow}), similar to the gap found by \citet{fultonetal2017}. However, further analysis with a larger planet sample is required to confirm the radius gap for \Ktwo\ planets.

This work has clear broader implications. The ability to validate planet candidates is vital to conducting a successful follow-up and confirmation program. The wide applicability of the validation infrastructure developed in this research is clear from the large number of candidates subjected to our validation process. By continuing to apply the validation process described here, future \Ktwo\ campaigns and the upcoming $TESS$ mission will benefit from a valuable source of validated planets and a useful validation pipeline able to process the large and constant supply of identified planet candidates.

\acknowledgments
Many thanks to Guillermo Torres and Dimitar Sasselov for their review and grading of this work during its phase as a thesis project. We thank Joey Rodriguez, George Zhou, Sam Quinn, Jeff Coughlin, and Tom Barclay for useful conversations and assistance in vetting some of our candidates. We are grateful to B.J. Fulton for providing us access to useful contour plotting tools, and also for allowing for the reproduction of a plot from another paper. Special thanks to Ellen Price for crucial assistance, time, and effort in setting up \vespa. Finally, A.W.M. wishes to thank Prof. David Charbonneau and the classmates of Astronomy 99 for their regular support and feedback.

A.W.M., A.V., and E.J.G. are supported by the NSF Graduate Research Fellowship grant nos. DGE 1752814, 1144152, and 1339067, respectively. This work was performed in part under contract with the California Institute of Technology (Caltech)/Jet Propulsion Laboratory (JPL) funded by NASA through the Sagan Fellowship Program executed by the NASA Exoplanet Science Institute. A.V. and D.W.L. acknowledge partial support from the $TESS$ mission through a sub-award from the Massachusetts Institute of Technology to the Smithsonian Astrophysical Observatory.

Some observations in the paper made use of the NN-EXPLORE Exoplanet and Stellar Speckle Imager (NESSI). NESSI was funded by the NASA Exoplanet Exploration Program and the NASA Ames Research Center. NESSI was built at the Ames Research Center by Steve B. Howell, Nic Scott, Elliott P. Horch, and Emmett Quigley. The NESSI data were obtained at the WIYN Observatory from telescope time allocated to NN-EXPLORE through the scientific partnership of the National Aeronautics and Space Administration, the National Science Foundation, and the National Optical Astronomy Observatory.

This paper includes data collected by the \Kepler/\Ktwo\ mission. Funding for the \Kepler\ mission is provided by the NASA Science Mission directorate. Some of the data presented in this paper were obtained from the Mikulski Archive for Space Telescopes (MAST). STScI is operated by the Association of Universities for Research in Astronomy, Inc., under NASA contract NAS5--26555. Support for MAST for non--HST data is provided by the NASA Office of Space Science via grant NNX13AC07G and by other grants and contracts. Support from the Kepler Participating Scientist Program was provided via NASA grant NNX14AE11G. The authors are honored to be permitted to conduct observations on Iolkam Du’ag (Kitt Peak), a mountain within the Tohono O'odham Nation with particular significance to the Tohono O'odham people. Some of the data presented herein were obtained at the WM Keck Observatory (which is operated as a scientific partnership among Caltech, UC, and NASA). The authors wish to recognize and acknowledge the very significant cultural role and reverence that the summit of Mauna Kea has always had within the indigenous Hawaiian community. We are most fortunate to have the opportunity to conduct observations from this mountain.

Facilities: \facility{Kepler, FLWO:1.5m (TRES), WIYN (DSSI, NESSI), Gemini:Gillett (DSSI, NIRI), Gemini:South (DSSI), Keck:II (NIRC2), Hale (PHARO), LBT (LMIRCam), Gaia, $HIPPARCOS$, Exoplanet Archive, ADS, MAST}

\appendix
%\begin{appendices}
\section{Calibrating TRES to Measure the Mt. Wilson $S_{HK}$ Index}

Our work involved the analysis of many spectra collected with TRES on the 1.5 m Tillinghast telescope at the Whipple Observatory on Mt. Hopkins. In addition to using TRES spectra to derive spectroscopic parameters and measure radial velocities, we also used TRES to measure the Mt. Wilson $S_{HK}$ activity indicator. $S_{HK}$ is a ratio between the flux in the cores of the calcium II $H$ and $K$ spectral features (at $3933.66 \pm 1.09$ and $3968.47 \pm 1.09$ \AA) and the flux in two nearby continuum regions (one slightly redward of the Ca II lines called $R$, and one slightly blueward called $V$). $S_{HK}$ is commonly used as a proxy for a star's chromospheric activity \citep{isaacsonandfischer2010}. Because $S_{HK}$ depends on ratios of fluxes at different wavelengths, it is sensitively affected by the spectrograph blaze function. The effect of the spectrograph blaze function must therefore be carefully calibrated and quantified in order to standardize measurements of $S_{HK}$ from different instruments. In this appendix, we describe how we calibrated the TRES spectrograph to place measurements of $S_{HK}$ on the Mt. Wilson scale, and present our measurements of $S_{HK}$ for a handful of bright K2 planet candidate hosts for which we were able to measure $S_{HK}$ reliably. These measurements are listed in Table~\ref{table:shk_table}.
% older version:
% and must be calibrated in order to be compared between  Calibration is required because the blaze function of the TRES spectrograph is unique, so the effect of the blaze function on the Ca H and K features must be quantified so that it can be removed. This calibration was applied to the TRES spectra analyzed in the body of this work (see Section~\ref{stellar_parameters}) and the $S_{HK}$ results can be found in Table~\ref{table:shk_table}.

To calibrate TRES to measure $S_{HK}$ on the Mt. Wilson scale, we closely followed the procedure of \citet{isaacsonandfischer2010}. We based our calibration on a sample of stars that were both observed by TRES (prior to 2016 June 1) and were included in the \citet{duncanetal1991} catalog of Mt. Wilson $S_{HK}$ measurements. Additionally, the following cuts were applied to improve the sample quality:

\begin{enumerate}
    \item TRES spectra for the star in question were only used if the photon count in the $R$ continuum region ($4001.07 \pm 10$ \AA) was greater than 174,000. (The cutoff is actually at 150,000 ADU, which is converted using a gain of $1.16$ for TRES.) This conservative cutoff was imposed to remove spectra with only low or moderate S/N from the calibration.
    \item The star was not in a close binary or multiple system. We only included widely separated ($\gtrsim 5$\arcsec) visual binaries in order to prevent flux from the companion star(s) being blended with flux of the intended target.
    \item One fast-rotating star (HD 30780, or `VB123' in the TRES database) was removed from the calibration set as the H and K emission cores were clearly broadened and smeared with other spectral features ($v\sin{i} \simeq $ 180 km s$^{-1}$).
\end{enumerate}

After these cuts, our calibration sample included 118 stars with a total of 1204 individual TRES observations. The calibration was made in the same way as in \citet{isaacsonandfischer2010}, by determining values of $K_{\rm coeff}$, $V_{\rm coeff}$, $m$, and $b$ that minimize the differences between the Mt. Wilson survey measurement of $S_{HK}$ for each star \citep{duncanetal1991} and $S_{HK}$ calculated from the TRES spectra using the following equation:

\begin{equation}\label{shkeqn}
    S_{\rm HK, TRES} = m \times \frac{H + K_{\rm coeff} K}{R+V_{\rm coeff} V} + b
\end{equation}

\noindent where $H$ and $K$ are fluxes in the cores of the Ca II $H$ and $K$ lines,\footnote{The $H$ and $K$ fluxes were calculated by integrating the spectrum over triangular-shaped bandpasses centered at 3933.66 and 3968.47 \AA, respectively, each with a FWHM of 1.09 \AA.} $R$ and $V$ are continuum fluxes \footnote{The $R$ and $V$ fluxes were calculated by integrating the spectrum over 20 \AA\ wide box-shaped bandpasses centered at $4001.07$ and $3901.07$ \AA, respectively, following \citet{duncanetal1991}.}, $K_{\rm coeff}$ is a coefficient that scales the magnitude of $K$ fluxes to match the magnitude of $H$ fluxes, $V_{\rm coeff}$ is a coefficient that scales the magnitude of $V$ fluxes to match the magnitude of $R$ fluxes, and $m$ and $b$ are free parameters.%  .

$K_{\rm coeff}$ was calculated through a linear regression between $H$ and $K$ for our TRES spectra, so as to scale the magnitude of $K$ to match that of $H$. The same process was performed for $R$ and $V$ in order to calculate $V_{\rm coeff}$. This was done to mirror the approach taken by \citet{isaacsonandfischer2010} in their calibration of the Keck HIRES instrument.

We estimated $m$ and $b$ by minimizing the differences between $S_{\rm HK, Mt. Wilson}$ from the Mt. Wilson survey \citep{duncanetal1991} and $S_{\rm HK, TRES}$ determined from TRES spectra with Equation \ref{shkeqn} using an MCMC algorithm with affine-invariant ensemble sampling \citep{foreman-mackeyetal2013}. We calculated the likelihood, $\mathcal{L}$, using the following function: % Also included in the model was a noise (or jitter) parameter, $\ln$(jitter), which we included in our likelihood function to compensate for a lack of error bars in the regression and account for, among other things, stellar activity variability between Mt. Wilson $S_{HK}$ measurements and TRES $S_{HK}$ measurements.

\begin{equation}
    \ln({\mathcal{L}}) = -0.5\sum\bigg(\frac{(S_{\rm HK, TRES} - S_{\rm HK, Mt. Wilson})^2}{\sigma^2} + \ln{\sigma^2}\bigg)
\end{equation}

\noindent where $\sigma$ is a jitter term that absorbs uncertainties in both in measurements of $S_{HK}$ and in their uncertainties due to factors including astrophysical variability in the $S_{HK}$ index of a single star over time. We imposed uniform priors on $m$ and $b$, and imposed a Jeffreys prior on $\sigma$. 

%\begin{equation}
%\sigma = e^{\ln(jitter)}
%\end{equation}

%Specifically, $\ln$(jitter) was included in the $\ln$(likelihood) function as follows:
%$$\sigma = e^{\ln(jitter)}$$
%$$\ln(likelihood) = -0.5\sum\bigg(\frac{(obs-model)^2}{\sigma^2} + \ln{\sigma^2}\bigg)$$
%Here $obs$ is all of the $S_{HK}$ observations made by Mt. Wilson and reported in \citet{duncanetal1991} and $model$ is all of the calculated $S_{HK}$ observations for TRES spectra using the equation for $S_{HK}$ above.

It is important to note that there might be both multiple TRES spectra and multiple Mt. Wilson $S_{HK}$ measurements for a given star. So if a star had more than one TRES spectrum or Mt. Wilson $S_{HK}$ measurement, this star was simply represented as a grid of TRES and Mt. Wilson measurements in the MCMC process. With this in mind, the total number of data points in the fitting process was 3108. This implicitly gives higher weight in our calibration to stars with multiple observations, which helps average out astrophysical variability in $S_{HK}$. 

\begin{figure}[hbtp]
\centering
\includegraphics[scale=0.4]{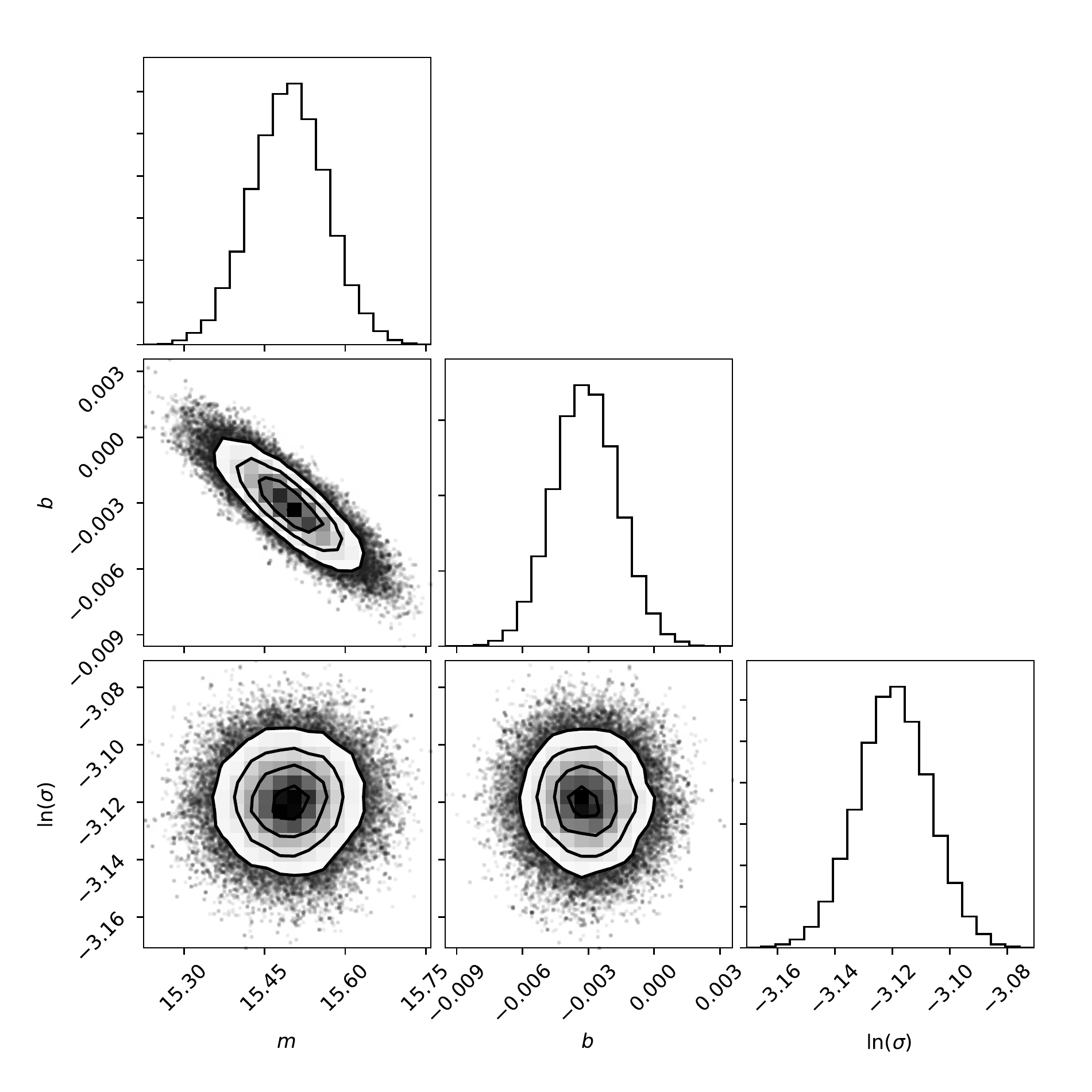}
\caption{Corner plot for three parameters in our calibration of the Tillinghast Reflector Echelle Spectrograph for $S_{HK}$, an indicator of stellar activity: a line slope ($m$) and a line $y$-intercept ($b$) in our $S_{HK}$ equation, and a noise parameter ($\ln{(\sigma)}$) to account for error bars and stellar variability.} \label{corner_plot}
\end{figure}

Our MCMC process used 200 chains of 50,000 steps each in order to explore parameter space and determine best-fit parameters and uncertainties. The resulting ensemble was well converged according to the Gelman-Rubin statistic \citep{gelmanandrubin1992}, which in this case was lower than 1.04 for all parameters. The last 500 samples of the MCMC process are visualized in Figure \ref{corner_plot} with a posterior corner plot \citep{foreman-mackey2016}. We found the following parameter values from our model fit:

$$K_{\rm coeff} = 0.876$$
$$V_{\rm coeff} = 0.775$$
$$m = 15.496^{+0.068}_{-0.069}$$
$$b = -0.0031 \pm 0.0015$$
$$\ln(\mathrm{jitter}) = -3.119 \pm 0.013$$

There was also significant covariance between $m$ and $b$ (see Fig.~\ref{corner_plot}), which we estimated to be $\mathrm{Cov}(m,b) = -8.571 \times 10^{-5}$.

Statistical uncertainties, $\sigma_{\rm random}$, for $S_{HK}$ were calculated by propagating sample draws from four Poisson distributions representing $R$, $V$, $H$, and $K$ through draws from the $m$ and $b$ posterior distributions and the $K_{\rm coeff}$ and $V_{\rm coeff}$ values. We also included a systematic error term, $\sigma_{\rm systematic}$, in our uncertainty estimates. We estimated the systematic uncertainties for TRES $S_{HK}$ measurements using six spectra with high S/N of the bright star Tau Ceti collected with TRES on the night of 2013 October 24. We assumed that $S_{HK}$ for Tau Ceti remained constant over the course of a single night, and estimated a systematic error term based on the scatter between individual $S_{HK}$ measurements. Our final error estimates, $\sigma_{\rm total}$ came from assuming that the statistical uncertainties and systematic uncertainties were independent and could be added in quadrature as follows:

\begin{equation}
    \sigma_{\rm total}^2 = \sigma_{\rm systematic}^2 + \sigma_{\rm random}^2
\end{equation}
%where total error was the standard deviation between the $S_{HK}$ values of the 6 spectra ($\simeq 0.0025$) and random error was the average of the photon-limited errors for each spectrum ($\simeq 0.0010$), yielding a systematic error estimate of $\simeq 0.0022$. This estimate of systematic error was then built into the calibration by adding it in quadrature with the photon-limited random error for the spectrum in question; the resulting total error was used as the error estimate for the $S_{HK}$ value.

% $S_{HK}$ values and corresponding photon-limited errors were calculated for these spectra. Then, an estimate of the systematic error of TRES was calculated by solving the following equation:

We note that our calibration for $S_{HK}$ is only valid over specific ranges in parameter values. Our calibration was performed over a range of $0.244-1.629$ in B-V and yielded $S_{HK}$ values ranging from $0.055$ to $2.070$. Based on tests with simulated data and with real data with low S/N, we are cautious of $S_{HK}$ measurements on spectra with fewer than 290,000 photon counts (or 250,000 ADU) in the $R$ and $V$ continuum regions combined. We also note that $S_{HK}$ measurements can be affected by rapid stellar rotation. We found by artificially broadening the lines of slowly rotating stars that $S_{HK}$ measurements for stars with $v\sin{i}$ above 20 km s$^{-1}$ can be skewed compared to the values that would have been measured if the stars were rotating slowly. % I wouldn't quote these numbers - I'm suspicious the poisson statistics tests don't fully capture the errors that can appear, and there are too many variables in the rotation tests to go into here. 

%star , and based on tests with rapidly rotating spectra, we found that  By artificially inflating Poisson noise, we found that a photon count of less than 250,000 in the R and V continuum regions could skew $S_{HK}$ by ${\sim}2\%$. Similarly, we found that artificially inflating $v\sin{i}$ above 20 km s$^{-1}$ could skew $S_{HK}$ by ${\sim}2\%$.

We tested our calibration by comparing $S_{HK}$ measurements derived from TRES with a test set of $S_{HK}$ measurements from the literature \citep[in particular, we focused on stars observed by ][]{isaacsonandfischer2010}.  We constructed the test data set by identifying stars observed by TRES with the following properties: % our The validity of the calibration was determined by comparing its performance on our training data (the 1204 TRES spectra for 118 stars and their corresponding Mt. Wilson $S_{HK}$ measurements) against a test data set.

\begin{enumerate}
    \item The stars were not included in our calibration set (stars observed both by \citet{duncanetal1991} and TRES).
    \item The TRES spectra for the star in question have photon counts in the $R$ continuum region greater than 174,000 (or 150,000 ADU).
    \item The star had at least one $S_{HK}$ measurement from Keck HIRES that had been reported in \citet{isaacsonandfischer2010}.
\end{enumerate}

After applying these cuts, we were left with a test set of 121 stars, which had been observed a total of 4308 times by TRES.

\begin{figure}[hbtp]
\centering
\includegraphics[scale=0.5]{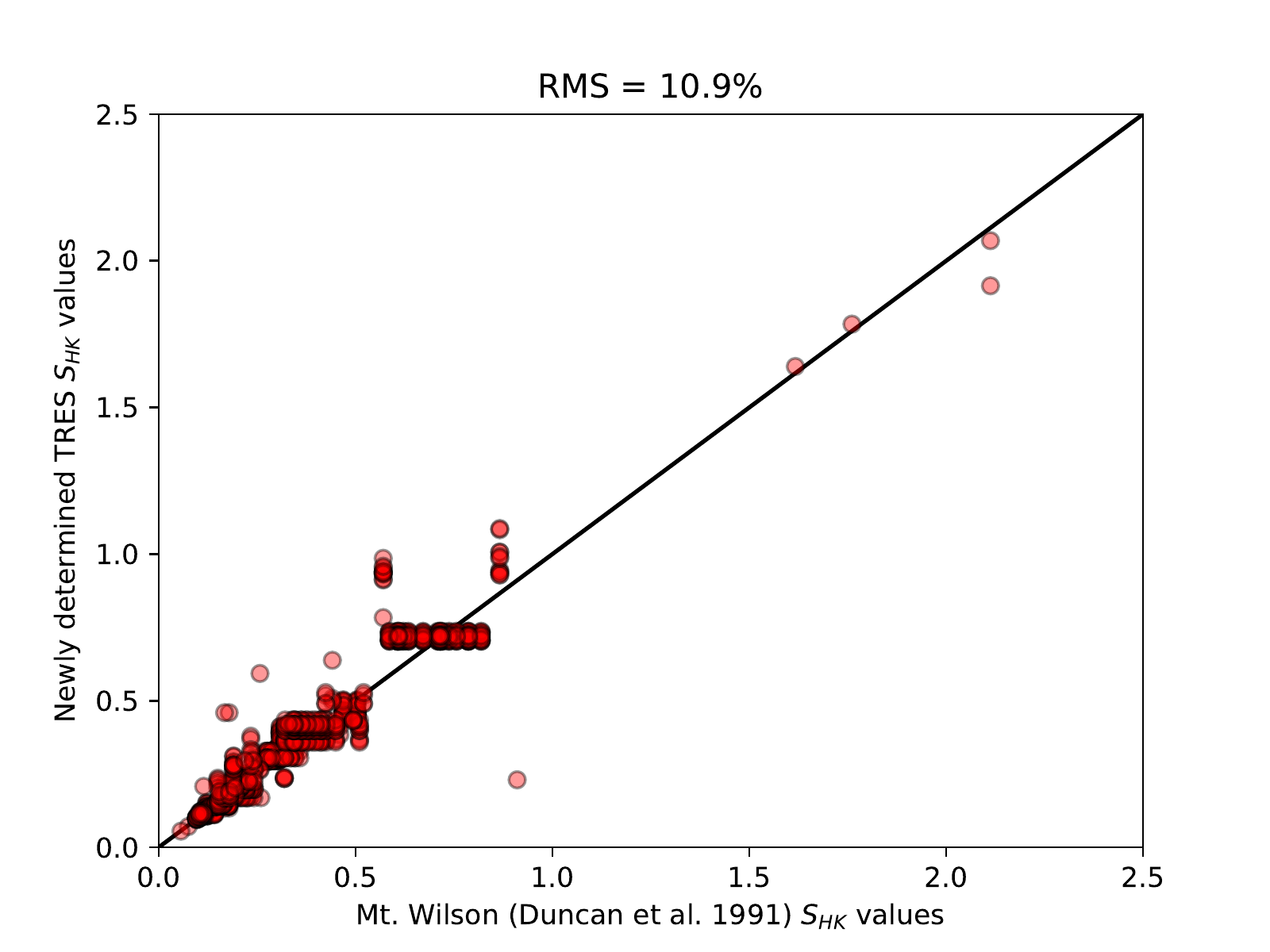}
\caption{Comparison of our $S_{HK}$ calibration for TRES with the \citet{duncanetal1991} catalog for our calibration data set of 118 stars (with 1204 TRES spectra). Each red point denotes a unique pair of observations from TRES and Mt. Wilson for a given star. The fractional RMS scatter on the calibration set is $11.0\%$.} \label{training_fit}
\end{figure}

\begin{figure}[hbtp]
\centering
\includegraphics[scale=0.5]{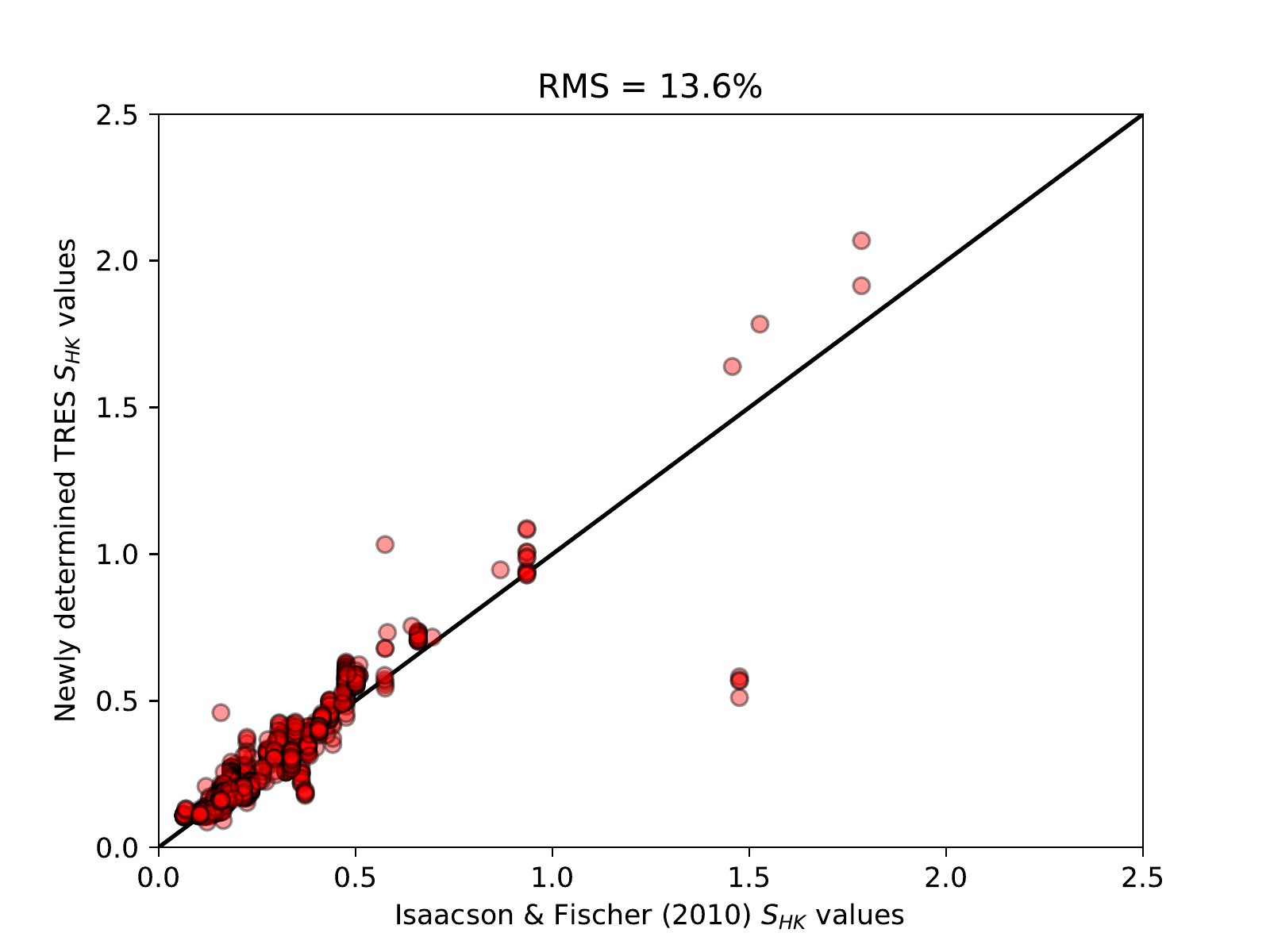}
\caption{Comparison of our $S_{HK}$ calibration for TRES with the \citet{isaacsonandfischer2010} catalog for our test data set of 121 stars (with 4308 TRES spectra). Each red point denotes a unique pair of observations from TRES and Mt. Wilson for a given star. The fractional RMS scatter on the calibration set is $13.6\%$.} \label{testing_fit}
\end{figure}

We show the results of our $S_{HK}$ calibration in Figure \ref{training_fit}, which compares TRES $S_{HK}$ measurements from our calibration set with measurements from the Mt. Wilson survey, and in Figure \ref{testing_fit}, which compares TRES $S_{HK}$ measurements from our test set with values from \citet{isaacsonandfischer2010}. % and testing sets, and the results for both data sets can be seen in Fig.~\ref{training_fit} and Fig.~\ref{testing_fit} respectively.

For the calibration set, there is an $10.9\%$ fractional RMS scatter about the one-to-one line. The only noticeable outlier appears to be at approximately (0.9,0.2), corresponding to the single TRES spectrum (and Mt. Wilson measurement) of HD 120477, a variable star observed in 2015 July by TRES and observed no later than 1991 at Mt. Wilson. The reason for this outlier could not be identified, but its presence in the calibration set did not significantly change the calibration, so it was left in (removing the point maintained a fractional RMS scatter of $10.9\%$.)

As for the test set, there is a $13.6\%$ fractional RMS scatter about the one-to-one line, slightly larger than for the calibration set (which is to be expected). There does appear to a slight systematic offset between TRES and Keck HIRES such that $S_{HK}$ values from Keck HIRES tend to be lower than those from TRES. However, this is to be expected, since the same systematic underestimation with Keck HIRES compared to Mt. Wilson can be seen in Fig.~4 of \citet{isaacsonandfischer2010}. There are also a few obviously errant data points in this figure, centered around (1.5, 0.5). These data points correspond to four TRES spectra of the star GJ 570B, a spectroscopic binary. The test data set, because it was not used for calibration, was not clipped of binary stars; thus, GJ 570B is a clear example of how calculating $S_{HK}$ for close binaries can go wrong. However, the presence of those four spectra has very little effect on the results; removing them only reduced the fractional RMS scatter of the test set to $13.5\%$.

%\tabletypesize{\tiny} 
\fontsize{9}{11}\selectfont
\LongTables
%\begin{small}
%\begin{table}[h] 
%\centering
\begin{deluxetable}{ccc}
\tabletypesize{\footnotesize} 
\tablecolumns{13} 
\tablewidth{0pt} 
\tablecaption{Stellar $S_{HK}$ Activity Index \label{table:shk_table}} 
\tablehead{EPIC & Date & $S_{HK}$}
\startdata 
201437844 & 2017 Jan 08 & $0.1666^{+0.0030}_{-0.0031}$ \\
201437844 & 2017 Mar 09 & $0.1698 \pm 0.0032$ \\
201437844 & 2017 Mar 09 & $0.1589 \pm 0.0031$ \\
201437844 & 2017 Mar 09 & $0.1712 \pm 0.0031$ \\
201437844 & 2017 Mar 09 & $0.1651^{+0.0030}_{-0.0029}$ \\
201437844 & 2017 Mar 09 & $0.1624 \pm 0.0030$ \\
201437844 & 2017 Mar 09 & $0.1633^{+0.0030}_{-0.0029}$ \\
201437844 & 2017 Mar 09 & $0.1614 \pm 0.0029$ \\
201437844 & 2017 Mar 09 & $0.1692 \pm 0.0029$ \\
201437844 & 2017 Mar 09 & $0.1622^{+0.0030}_{-0.0029}$ \\
201437844 & 2017 Mar 09 & $0.1635 \pm 0.0028$ \\
201437844 & 2017 Mar 09 & $0.1593^{+0.0029}_{-0.0027}$ \\
201437844 & 2017 Mar 09 & $0.1615^{+0.0028}_{-0.0029}$ \\
201437844 & 2017 Mar 09 & $0.1680^{+0.0028}_{-0.0029}$ \\
201437844 & 2017 Mar 09 & $0.1646 \pm 0.0029$ \\
201437844 & 2017 Mar 09 & $0.1639 \pm 0.0028$ \\
201437844 & 2017 Mar 09 & $0.1651 \pm 0.0028$ \\
201437844 & 2017 Mar 09 & $0.1656 \pm 0.0029$ \\
201437844 & 2017 Mar 09 & $0.1693 \pm 0.0029$ \\
201437844 & 2017 Mar 09 & $0.1641 \pm 0.0029$ \\
201437844 & 2017 Mar 09 & $0.1680 \pm 0.0029$ \\
205904628 & 2015 Jul 24 & $0.1550^{+0.0029}_{-0.0030}$ \\
205904628 & 2015 Sep 06 & $0.1453^{+0.0031}_{-0.0033}$ \\
211424769 & 2015 Dec 09 & $0.3255^{+0.0040}_{-0.0039}$ \\
211424769 & 2016 Jan 01 & $0.3144^{+0.0040}_{-0.0039}$ \\
211993818 & 2015 Nov 27 & $0.4189^{+0.0044}_{-0.0045}$ \\
211993818 & 2016 Feb 29 & $0.3851^{+0.0031}_{-0.0030}$ \\
211993818 & 2016 Mar 02 & $0.3115^{+0.0027}_{-0.0028}$
\enddata
%\tablecomments{} 
\end{deluxetable}
%\end{small}

\clearpage
%\end{appendices}

\bibliographystyle{apj}
\bibliography{refs}

% \documentclass[iop,revtex4]{emulateapj}

% \usepackage{makecell}
% \usepackage{hhline}

% \usepackage{lscape}
% \usepackage{amsmath}

% \begin{document}

\tabletypesize{\tiny} 
\clearpage % optional
\fontsize{9}{11}\selectfont
\LongTables
\begin{small}
\begin{landscape}
\begin{deluxetable}{>{\centering}m{1.3cm}ccccccccccccc}
\tabletypesize{\footnotesize} 
\tablecolumns{14} 
\tablewidth{0pt} 
\tablecaption{Planet Candidate Parameters \label{table:candidate_parameters_table}} 
\tablehead{ 
\Ktwo\ Name & EPIC & $T_0$ (BJD-2454833) & $P$ (days) & $a/R_*$ & $i$ (deg) & $R_\mathrm{p}/R_*$ & $R_\mathrm{p}$ ($R_\oplus$) & $R_*$ ($R_\odot$) & $M_*$ ($M_\odot$) & $K_\mathrm{p}$ & FPP & Disposition & Notes}
\startdata 
K2-156 b & 201110617.01 & $2750.1409 \pm 0.0028$ & $0.813149^{+0.000050}_{-0.000049}$ & $4.4^{+0.7}_{-1.1}$ & $84.0^{+4.3}_{-7.6}$ & $0.0170^{+0.0014}_{-0.0011}$ & $1.14^{+0.10}_{-0.08}$ & $0.616^{+0.019}_{-0.016}$ & $0.642^{+0.022}_{-0.019}$ & $12.947$ & $<1.00e-04$ & Planet &  \\
 & 201111557.01 & $2750.1621 \pm 0.0021$ & $2.30237^{+0.00011}_{-0.00010}$ & $12.6^{+2.8}_{-8.0}$ & $87.4^{+1.9}_{-9.5}$ & $0.0169^{+0.0067}_{-0.0015}$ & $1.31^{+0.52}_{-0.12}$ & $0.711^{+0.019}_{-0.020}$ & $0.746 \pm 0.023$ & $11.363$ & $1.22e-03$ & Candidate &  \\
 & 201127519.01 & $2752.5513^{+0.0011}_{-0.0012}$ & $6.17837^{+0.00019}_{-0.00017}$ & $17.3^{+1.1}_{-2.0}$ & $88.8^{+0.8}_{-1.0}$ & $0.1151^{+0.0049}_{-0.0034}$ & $9.91^{+0.53}_{-0.36}$ & $0.789^{+0.025}_{-0.017}$ & $0.857^{+0.021}_{-0.023}$ & $11.558$ & $ - $ & Candidate & c \\
K2-157 b & 201130233.01 & $2749.8239^{+0.0032}_{-0.0033}$ & $0.365257 \pm 0.000029$ & $1.88^{+0.32}_{-0.38}$ & $77.0^{+10.0}_{-18.0}$ & $0.0110^{+0.0014}_{-0.0010}$ & $1.06^{+0.15}_{-0.10}$ & $0.876^{+0.045}_{-0.025}$ & $0.940^{+0.023}_{-0.027}$ & $12.604$ & $<1.00e-04$ & Planet &  \\
K2-158 b & 201132684.01 & $2757.4817^{+0.0096}_{-0.0092}$ & $10.0621^{+0.0023}_{-0.0022}$ & $17.6^{+2.9}_{-5.8}$ & $88.4^{+1.2}_{-2.4}$ & $0.0271^{+0.0027}_{-0.0020}$ & $2.63^{+0.35}_{-0.22}$ & $0.892^{+0.074}_{-0.036}$ & $0.933^{+0.027}_{-0.030}$ & $11.678$ & $<1.00e-04$ & Planet &  \\
 & 201166680.01 & $2765.193^{+0.021}_{-0.016}$ & $18.105^{+0.010}_{-0.013}$ & $26.3^{+5.3}_{-8.9}$ & $88.9^{+0.8}_{-1.7}$ & $0.0157^{+0.0017}_{-0.0012}$ & $2.10^{+0.31}_{-0.20}$ & $1.22^{+0.12}_{-0.07}$ & $1.192^{+0.033}_{-0.030}$ & $10.897$ & $1.73e-03$ & Candidate &  \\
 & 201211526.01 & $2755.4780 \pm 0.0040$ & $21.0702^{+0.0024}_{-0.0023}$ & $34.0^{+5.0}_{-11.0}$ & $89.2^{+0.6}_{-1.2}$ & $0.0170^{+0.0031}_{-0.0013}$ & $1.65^{+0.32}_{-0.14}$ & $0.891^{+0.056}_{-0.037}$ & $0.936^{+0.032}_{-0.031}$ & $11.696$ & $2.10e-03$ & Candidate &  \\
\enddata
\tablenotetext{a}{Unless there is a deep secondary eclipse or ellipsoidal variations in the light curve, \vespa\ cannot distinguish between a planet-sized star and a planet. In this case, $R_\mathrm{p} > 8 R_\oplus$ and RV measurements cannot rule out the foreground eclipsing binary scenario. Therefore, no FPP value is reported.}
\tablenotetext{b}{Companion in aperture, therefore no FPP is reported}
\tablenotetext{c}{AO/Speckle companion, therefore no FPP is reported}
\tablenotetext{d}{Composite spectrum, therefore no FPP is reported}
\tablenotetext{e}{Large RV amplitude variations confirm a binary in the system, therefore no FPP is reported}
\tablenotetext{f}{\vespa\ failed to find a FPP}
\tablenotetext{g}{HD 106315}
\tablenotetext{h}{K2-51 is a false positive (see \citet{shporeretal2017} and Section \ref{full_results})}
\tablenotetext{i}{WASP-152}
\tablenotetext{j}{WASP-157}
\tablenotetext{k}{HATS-12}
\tablenotetext{l}{HD 3167}

\end{deluxetable}
\clearpage
\end{landscape}
\end{small}

% \end{document}

% \documentclass[iop]{emulateapj}

% \begin{document}

\tabletypesize{\small} 
\clearpage % optional
\fontsize{9}{11}\selectfont
\LongTables
\begin{small}
%\begin{table}[h] 
%\centering
\begin{deluxetable}{cccccccc}
\tablecolumns{8} 
\tablewidth{0pt}
\tablecaption{Stellar Parameters
\label{table:stellar_parameters_table}} 
\tablehead{ 
EPIC & $T_{\mathrm{eff}}$ & [$m/\mathrm{H}$] & $\log{g}$ & $R_*$ ($R_\odot$) & $M_*$ ($M_\odot$) & $K_\mathrm{p}$ & Notes}

\startdata 
201110617 & $4460.0 \pm 50.0$ & $-0.33 \pm 0.08$ & $4.68 \pm 0.1$ & $0.616^{+0.019}_{-0.016}$ & $0.642^{+0.022}_{-0.019}$ & $12.947$ &  \\
201111557 & $4720.0 \pm 50.0$ & $-0.06 \pm 0.08$ & $4.5 \pm 0.1$ & $0.711^{+0.019}_{-0.020}$ & $0.746 \pm 0.023$ & $11.363$ &  \\
201127519 & $5015.0 \pm 50.0$ & $0.24 \pm 0.08$ & $4.67 \pm 0.1$ & $0.789^{+0.025}_{-0.017}$ & $0.857^{+0.021}_{-0.023}$ & $11.558$ &  \\
201130233 & $5456.0 \pm 50.0$ & $0.13 \pm 0.08$ & $4.55 \pm 0.1$ & $0.876^{+0.045}_{-0.025}$ & $0.940^{+0.023}_{-0.027}$ & $12.604$ &  \\
201132684 & $5489.0 \pm 50.0$ & $0.07 \pm 0.08$ & $4.46 \pm 0.1$ & $0.892^{+0.074}_{-0.036}$ & $0.933^{+0.027}_{-0.030}$ & $11.678$ &  \\
201166680 & $6213.0 \pm 50.0$ & $0.14 \pm 0.08$ & $4.32 \pm 0.1$ & $1.22^{+0.12}_{-0.07}$ & $1.192^{+0.033}_{-0.030}$ & $10.897$ &  \\
201211526 & $5728.0 \pm 50.0$ & $-0.2 \pm 0.08$ & $4.51 \pm 0.1$ & $0.891^{+0.056}_{-0.037}$ & $0.936^{+0.032}_{-0.031}$ & $11.696$ &  \\
201225286 & $5419.0 \pm 50.0$ & $-0.1 \pm 0.08$ & $4.57 \pm 0.1$ & $0.830^{+0.045}_{-0.029}$ & $0.886^{+0.027}_{-0.032}$ & $11.729$ &  \\
\enddata
\tablenotetext{a}{Stellar parameters for this star are derived from a single spectrum with a cross-correlation function peak height $< 0.9$ (but $> 0.8$). We have decided that the resulting stellar parameters are trustworthy for this analysis, but provide a note for the reader's discretion.}
\tablenotetext{b}{$S_{HK}$ values have been determined for this star and can be found in Table~\ref{table:shk_table}.}
\end{deluxetable}
\clearpage % optional
\end{small}

% \end{document}
% \documentclass[iop,revtex4]{emulateapj}

% \usepackage{makecell}
% \usepackage{hhline}

% \usepackage{lscape}
% \usepackage{amsmath}

% \begin{document}

\tabletypesize{\tiny} 
\tablewidth{0pt}
\fontsize{9}{11}\selectfont
\LongTables
\begin{tiny}
\begin{landscape}
\begin{deluxetable}{cccccccccccccccc}
\tablecolumns{16} 
\tablecaption{Detailed FPP Table
\label{table:fpp_table}} 
\tablehead{ 
\thead{\Ktwo\ Name} & \thead{EPIC} & \thead{Secondary\\Eclipse\\Threshold\\(ppm)} & \thead{Aperture\\Radius\\(arcsec)} & \thead{AO/\\Speckle} & \thead{\textit{P}(Eclipsing\\Binary)} & \thead{\textit{P}(Eclipsing\\Binary)\\(x2 Per.)} & \thead{\textit{P}(Back-\\ground\\Eclipsing\\Binary)} & \thead{\textit{P}(Back-\\ground\\Eclipsing\\Binary)\\(x2 Per.)} & \thead{\textit{P}(Hier-\\archical\\Eclipsing\\Binary)} & \thead{\textit{P}(Hier-\\archical\\Eclipsing\\Binary)\\(x2 Per.)} & \thead{RV\\Mass\\Limit} & \thead{VESPA\\FPP} & \thead{Multi-\\plicity\\Boost} & \thead{Adopted\\FPP} & \thead{Disposition}}

\startdata 
K2-156 b & 201110617.01 & $28$ & $14.7$ & Y & $0.00e+00$ & $0.00e+00$ & $<1e-04$ & $<1e-04$ & $<1e-04$ & $<1e-04$ & Y & $2.18e-08$ & ... & $<1e-04$ & Planet \\
 & 201111557.01 & $47$ & $14.4$ & Y & $<1e-04$ & $1.16e-03$ & $<1e-04$ & $<1e-04$ & $<1e-04$ & $<1e-04$ & N & $1.22e-03$ & ... & $1.22e-03$ & Candidate \\
 & 201127519.01 & $813$ & $17.9$ & Y & ... & ... & ... & ... & ... & ... & Y & ... & ... & ... & Candidate \\
K2-157 b & 201130233.01 & $5$ & $12.4$ & Y & $<1e-04$ & $<1e-04$ & $<1e-04$ & $<1e-04$ & $<1e-04$ & $<1e-04$ & N & $2.81e-08$ & ... & $<1e-04$ & Planet \\
K2-158 b & 201132684.01 & $95$ & $19.2$ & Y & $<1e-04$ & $<1e-04$ & $<1e-04$ & $<1e-04$ & $<1e-04$ & $<1e-04$ & N & $3.62e-06$ & ... & $<1e-04$ & Planet \\
 & 201166680.01 & $64$ & $17.9$ & Y & $1.70e-03$ & $<1e-04$ & $<1e-04$ & $<1e-04$ & $<1e-04$ & $<1e-04$ & N & $1.73e-03$ & ... & $1.73e-03$ & Candidate \\
 & 201211526.01 & $40$ & $17.3$ & Y & $<1e-04$ & $2.03e-03$ & $<1e-04$ & $<1e-04$ & $<1e-04$ & $<1e-04$ & N & $2.10e-03$ & ... & $2.10e-03$ & Candidate \\
K2-159 b & 201225286.01 & $85$ & $19.9$ & Y & $<1e-04$ & $3.43e-04$ & $<1e-04$ & $<1e-04$ & $<1e-04$ & $<1e-04$ & N & $3.53e-04$ & ... & $3.53e-04$ & Planet \\
\enddata
\tablenotetext{a}{HD 106315}
\tablenotetext{b}{K2-51 is a false positive (see \citet{shporeretal2017} and Section \ref{full_results})}
\tablenotetext{c}{WASP-152}
\tablenotetext{d}{WASP-157}
\tablenotetext{e}{HATS-12}
\tablenotetext{f}{HD 3167}
\end{deluxetable}
\clearpage
\end{landscape}
\end{tiny}

% \end{document}
% \documentclass[iop]{emulateapj}

% \begin{document}

\tabletypesize{\footnotesize} 
\clearpage % optional
\fontsize{9}{11}\selectfont
\LongTables
\begin{small}
\begin{deluxetable}{ccc}
\tablecolumns{8} 
\tablewidth{0pt}
\tablecaption{High-resolution Imaging
\label{table:ao_speckle_table}} 
\tablehead{ 
EPIC & Filter & Instrument}

\startdata 
201110617 & 562 & NESSI \\
201110617 & 832 & NESSI \\
201111557 & 562 & NESSI \\
201111557 & 832 & NESSI \\
201127519 & 562 & NESSI \\
201127519 & 832 & NESSI \\
201130233 & 562 & NESSI \\
201130233 & 832 & NESSI \\
\enddata
\end{deluxetable}
\clearpage % optional
\end{small}

\clearpage

\end{document}